\documentclass{article}

\usepackage{amsfonts}
\usepackage{amsmath}
\usepackage{epstopdf}
\usepackage{subfigure}
\usepackage{nomencl}
\makenomenclature
\usepackage{color}
\usepackage{url}
\usepackage{float}

\usepackage{tikz}
\usetikzlibrary{shapes.geometric, arrows} 
\usepackage{xcolor}
\usepackage[export]{adjustbox}
\usepackage{multicol}
\usepackage{gensymb}

\begin{document}



\title{Identifying Weakly Connected Subsystems in Building Energy Model for Effective Load Estimation in Presence of Parametric Uncertainty}

\author{Arpan Mukherjee\thanks{Corresponding author: Email: arpanmuk@buffalo.edu} \and Anna Kuechle Szweda, \and Andrew Alegria, \and Rahul Rai,\and Tarunraj Singh}
\date{Department of Mechanical and Aerospace Engineering, University at Buffalo-SUNY}

\maketitle

\begin{abstract}

It is necessary to estimate the expected energy usage of a building to determine how to reduce energy usage. The expected energy usage of a building can be reliably simulated using a Building Energy Model (BEM). Many of the numerous input parameters in a BEM are uncertain. To ensure that the building simulation is sufficiently accurate, and to better understand the impact of imprecisions in the input parameters and calculation methods, it is desirable to quantify uncertainty in the BEM throughout the modeling process. Uncertainty quantification (UQ) typically requires a large number of simulations to produce meaningful data, which, due to the vast number of input parameters and the dynamic nature of building simulation, is computationally expensive. Uncertainty Quantification (UQ) in BEM domain is thus intractable due to the size of the problem and parameters involved and hence it needs an advanced methodology for analysis. The current paper outlines a novel Weakly-Connected-Systems (WCSs) identification-based UQ framework developed to propagate the quantifiable uncertainty in the BEM. The overall approach is demonstrated on the physics-based thermal model of an actual building in Central New York. 
\end{abstract}

\textbf{keywords:} Building Energy Simulation, Uncertainty Quantification, Thermal Load Estimation, Weakly Connected Subsystem

\mbox{}

\nomenclature{$T_{amb}$}{Ambient temperature at exterior of surface $j$ ($\degree$C)}
\nomenclature{$T_k$}{Air temperature of zone $k$ ($\degree$C)}
\nomenclature{$T_o^j$}{Temperature of outside of surface $j$ ($\degree$C)}
\nomenclature{$T_i^j$}{Temperature of inside of surface $j$ ($\degree$C)}
\nomenclature{$A_{gross}^j$}{Gross area of surface $j$ (m$^2$)}
\nomenclature{$A^j$}{Net area of surface $j$ (m$^2$)}
\nomenclature{$M_k$}{Mass of air in zone k (kg)}
\nomenclature{$c_{pa}$}{Specific heat of air (J/kg$\cdot\degree$C)}
\nomenclature{$T_k^{ne}$}{Heat gain from non-envelope internal loads (W)}
\nomenclature{$Q_{structure}$}{Heat transfer between the structure (surface and windows) and zone $k$ air (W)}
\nomenclature{$h_o^j$}{Outside surface conductance of surface $j$ (W$\cdot$m$^2\cdot\degree$C)}
\nomenclature{$h_i^j$}{Inside surface conductance of surface $j$ (W$\cdot$m$^2\cdot\degree$C)}
\nomenclature{$C^j$}{Thermal capacitance of surface $j$ (J/m$\cdot\degree$C)}
\nomenclature{$R^j$}{Thermal resistance of surface $j$ ($\degree$C/W)}
\nomenclature{$A^w_{win}$}{Area of window $w$ (m$^2$)}
\nomenclature{$R^w_{win}$}{Thermal resistance of window $w$ ($\degree$C/W)}
\nomenclature{$U_o^j$}{Exterior solar heat gains on surface $j$ (W)}
\nomenclature{$T_q^j$}{Interior solar heat gains on surface $j$ (W)}
\nomenclature{$T^j_{out}$}{Temperature of the space adjacent to the surface $j$ ($\degree$C)}
\nomenclature{$\dot{m_k}$}{HVAC supply air mass flow rate for zone $k$ (kg/h)}
\nomenclature{$U_k^{sa}$}{Supply air temperature from the HVAC system for zone $k$ ($\degree$C)}
\nomenclature{$\alpha^j$}{Solar absorptance of surface $j$ (dimensionless)}  
\nomenclature{$I_{b}$}{Intensity of beam (direct) radiation (W$\cdot$degrees$^2$)}
\nomenclature{$A_s^j$}{Sunlit area of surface $j$ (m$^2$)}
\nomenclature{$\theta^j$}{Angle of incidence of the sun at surface $j$ (degrees)}
\nomenclature{$I_{s}$}{Intensity of sky diffuse radiation (W$\cdot$degrees$^2$)}
\nomenclature{$F_{ss}^j$}{Angle factor between surface $j$ and sky (dimensionless)}
\nomenclature{$F_{sg}^j$}{Angle factor between surface $j$ and ground (dimensionless)}
\nomenclature{$I_{g}$}{Intensity of ground reflected diffuse radiation (W$\cdot$degrees$^2$)}
\nomenclature{$Q^k_{isol}$}{Total heat gain through windows in zone $k$ (W)} 
\nomenclature{$SHGC^w$}{Solar Heat Gain Coefficient of window $w$ (dimensionless)}
\nomenclature{$A^w$}{Area of window $w$ (m$^2$)}
\nomenclature{$E_t^w$}{Incident total irradiance of window $w$ (W$\cdot$m$^2$)}
\nomenclature{$C_p^g$}{Specific heat of material $g$ (J/kg$\cdot\degree$C)}
\nomenclature{$l^g$}{Thickness of material $g$ (m)}
\nomenclature{$\rho^g$}{Density of material $g$ (kg/m$^3$)}
\nomenclature{$k^g$}{Thermal conductivity of material $g$ (W/m$\cdot\degree$C)}

\printnomenclature


\section{Introduction}
In 2016, the building sector used approximately 40\% of the energy produced in the United States~\cite{useia2017}, and in 2009, buildings contributed 2,184.6 million metric tons of carbon dioxide equivalent of greenhouse gases due to emissions~\cite{useia2009}.  As a result, governments and municipalities have committed to reducing the energy consumption of buildings.  New York State, for example, enacted Executive Order 88, which requires government buildings to reduce energy consumption by 20\% \cite{statedirect}, and the Public Service Commission in New York has established a public fund to target energy efficiency measures~\cite{nyspublic}. 

It is necessary to estimate the expected energy usage of a building to determine how to reduce energy usage. The expected energy usage of a building can be reliably simulated using a Building Energy Model (BEM). Currently, there are several popular options for software to estimate the dynamic energy use, such as EnergyPlus, eQUEST, TRACE 700, and Carrier HAP~\cite{Eplus,hirsch2006equest, trace, HAP}, each of which utilizes complex numerical methods to calculate building loads. The BEMs incorporate energy load estimation in the presence of HVAC systems, lights, and receptacle loads, and requires specification of a substantial number of input parameters.   

Many of the numerous input parameters in a BEM are uncertain. Some of these parameters are assumed in the model, while some of these parameters are measured. Additionally, the owner, designer, or modeling professional must estimate others. In each case, uncertainty is brought into the modeling results.  These uncertainties in the BEM can be categorized into two types: uncertainties in the measured parameters and the uncertainty in modeling~\cite{ding2015uncertainty,sun2015quantification}. To ensure that the building simulation is sufficiently accurate, and to better understand the impact of imprecisions in the input parameters and calculation methods, it is desirable to quantify uncertainty in the BEM throughout the modeling process. Uncertainty quantification (UQ) typically requires a large number of simulations to produce meaningful data, which, due to the vast number of input parameters and the dynamic nature of building simulation, is computationally expensive~\cite{rysanek2013optimum,eisenhower2012uncertainty}. We must, therefore, simplify the BEM to a less computationally expensive model.
To enable computationally efficient UQ of BEM, the following main contributions are made in this paper:

\begin{itemize}
\item Adaptation of lumped resistor-capacitance network model to generate reduced order differential equation for BEM simulation
\item Usage of a gray-box method in conjunction with black-box Kalman filter to enable BEM parameter estimation
\item Application of a novel WCSs identification-based UQ framework developed~\cite{Mukherjee_2017} to propagate the quantifiable uncertainty in the BEM.
\end{itemize}

Next, all the three main contributions are further elaborated. The well-established technique of a lumped resistor-capacitor (RC) network is used to model the reduced-order zone heat balance equations. The RC method assumes that the components of the zone load can be estimated by a discrete number of resistances and capacitances, and the system is treated as equivalent to an electrical circuit~\cite{vivian2017evaluation}. In particular, the heat transfer due to the zone envelope can be reduced to a three resistance, two capacitance (3R2C) thermal network. With the RC network, one can reduce the load calculations to first-order differential equations that provide a suitable framework to carry out UQ in BEMs. 

The BEM simulation can be carried out using one of the three main models~\cite{he2016simplified,perera2014modeling}:

\begin{enumerate}
\item \textit{White box models:} In white box models detailed information about the known physical process is used to predict the future states.
\item \textit{Black-box models:} In black-box models measured data is used to estimate non-physical parameters that abstractly represent the BEM performance.
\item \textit{Gray-box models:} Gray-box models combine both the white-box and black-box methods through the use of a reduced order model with known building properties for modeling the physical process. Such models can also be easily combined with Kalman filtering to enable parameter estimation. Usage of gray-box models reduces computation time while improving the accuracy of predictions.
\end{enumerate}

Because of a large number of input parameters, despite the reduced order model, the resulting filtering problem is computationally expensive. The novel method of identifying Weakly Connected Subsystems (WCSs)~\cite{mukherjee2015laplacian,mukherjee2015non,Mukherjee_2017} is used to address this problem of performing UQ in the BEM. WCS-based UQ approach allows us to group coupled parameters to minimize the associated computation time required while continuing to propagate the quantifiable uncertainty present in the BEM. The optimal estimation problem to estimate parameters for UQ of BEM is the heart of this research.

A representative case study, a well-documented building, located in Central New York, is used for modeling purposes. A separately simulated model of the represented building provides input data for the reduced order model, which we treat as actually measured information, with some initial uncertainty. 



The paper is organized as follows: Section~\ref{literature} describes the related works in two distinct but related domains: (1) the lumped capacitance RC network and (2) uncertainty quantification.  Section~\ref{method} explains the methodology followed to formulate the state space matrix and the associated input parameters for the model.  Section~\ref{method} also outlines the Weakly Connected Subsystems based solution methodology.  Section~\ref{case_study} provides details related to the case study building model.    Section~\ref{results} shows the results of the simulation, the cluster analysis, estimated temperatures, and non-envelope loads.  Section~\ref{conclusion} discusses the main conclusions and potential avenues for future work.

\section{Related Works}
\label{literature}
\subsection{Lumped capacitance RC network }
The lumped capacitance RC network reduced-order model method has been reliably used in a diverse set of research work to estimate the heat transfer due to the building envelope ~\cite{ramallo2013lumped,vsiroky2011experimental,li2017development}.  Further precision has been incorporated into the RC network by using dynamically estimated capacitances ~\cite{jara2016new} and second-order thermal network models ~\cite{underwood2014improved}.  The physics-based Three resistor-Two Capacitance (3R2C) model has been shown to be sufficiently accurate for approximating building models ~\cite{kircher2015lumped} based on parameters with physical meaning. In the current work, we build upon 3R2C models to enable our modeling framework.

The lumped capacitance method has been used to estimate other factors affecting the thermal zone heating and cooling requirements.  The thermal mass of the zone, in particular, can be modeled with an RC network ~\cite{perera2014modeling,wang2006parameter}.  When these values are significant, long-wave radiation ~\cite{he2016simplified} and convective heat between zones ~\cite{goyal2011identification} can be modeled as black-box data-driven estimations.  Internal loads, especially, have been estimated using a filtering method in conjunction with the RC method for the surfaces ~\cite{o2010model}.

The existing literature related to the RC method is inadequate for modeling large-scale problems due to high computational cost. This computational cost is exacerbated due to increase in the number of inputs. Thus, an essential contribution is in the development of a computational method to expedite the UQ using the idea of \textit{divide and conquer}. We outline such approach in this paper.

\subsection{Uncertainty Quantification in BEMs}

Uncertainty Quantification (UQ) is becoming more prevalent in the BEM domain, as building simulation software and methods continue to evolve~\cite{woloszyn2017treating}. Research in the field of UQ in BEM is growing at a fast pace. As explained earlier, uncertainty in a BEM can arise due to the modeling process, or uncertainty in the parameters. All BEMs necessarily make assumptions to simplify the modeled building. Simplifying the assumptions in the building energy simulation domain causes the associated uncertainty to be often ignored. Few existing works have focused on UQ in BEMs. For example, Sun et al.~\cite{sun2015quantification} have focused on incorporating the uncertainty in the solar irradiation calculation and its effects on the results of BEM simulation. Additionally, post-processing techniques have been developed for incorporating UQ in BEMs~\cite{ding2015uncertainty}. 

Recent studies have used parameter estimation for UQ in BEMs. Both physics-based and surrogate models have been studied to optimize building performance under uncertainties by simulation methods~\cite{nguyen2014review}. It has been shown that even moderately variable parameters can have a significant effect on the overall BEM uncertainty, especially, when the small-scale models are combined to generate a large-scale BEM, as demonstrated in multi-building residential district models~\cite{kavgic2015application,baetens2016modelling}.  Sensitivity analysis has often been used to determine the impact of uncertainty in parameters on the performance of the building energy model~\cite{rodriguez2013uncertainties,tian2013review}. Most of the work in sensitivity analysis focuses on identifying a few most influential parameters that can be used for further investigation during UQ. Additionally, most of the reported works are often limited to evaluating only particular aspects of a building and not the whole building~\cite{baetens2016modelling}.  

UQ in BEMs is also performed in Stochastic Model Predictive Control (SMPC) frameworks that utilize the dynamic state-space equations. Oldewurtel et al. have discussed how SMPC framework can be used to account for the uncertainty in weather predictions for building energy modeling purposes~\cite{oldewurtel2012use}. In another related work, an SPMC has been developed to optimize building energy usage and has been shown to outperform existing Rule-based Control (RBC)~\cite{oldewurtel2010energy} framework. SPMC has also been used for large-scale building (a large number of state variables) in the presence of uncertainties in weather prediction~\cite{oldewurtel2014stochastic}. Privara et al.~\cite{privara2011model} have used an identified state-space model of a real building to estimate the optimal energy consumption. The model developed by Privara et al. does not include the effect of internal loads or solar gains. An adaptive MPC has been developed incorporating uncertainties in a wide range of parameters including building materials, their thermal properties, and HVAC parameters~\cite{kim2013building}.

A simulation over an extensive range of values is required with a minimum computational expense to enable UQ.  To reduce the computation time, techniques such as quasi-random sampling ~\cite{eisenhower2012uncertainty} and pre-processing historical data in modeling predictive controls~\cite{maasoumy2014handling} have been used. Similarly, other sampling-based methods or quadrature-based methods can also be applied~\cite{stroud1971approximate}. When applied to large-scale problems with many input parameters, the performance of existing methods is computationally inefficient. We address these shortcomings in this paper through the use of Weakly Connected Subsystems based UQ method ~\cite{Mukherjee_2017,mukherjee2015non,mukherjee2015laplacian}.

\section{Methodology}
\label{method}

\subsection{Proposed Framework}

\begin{figure}[H]
\centering
\includegraphics[width=0.8\textwidth]{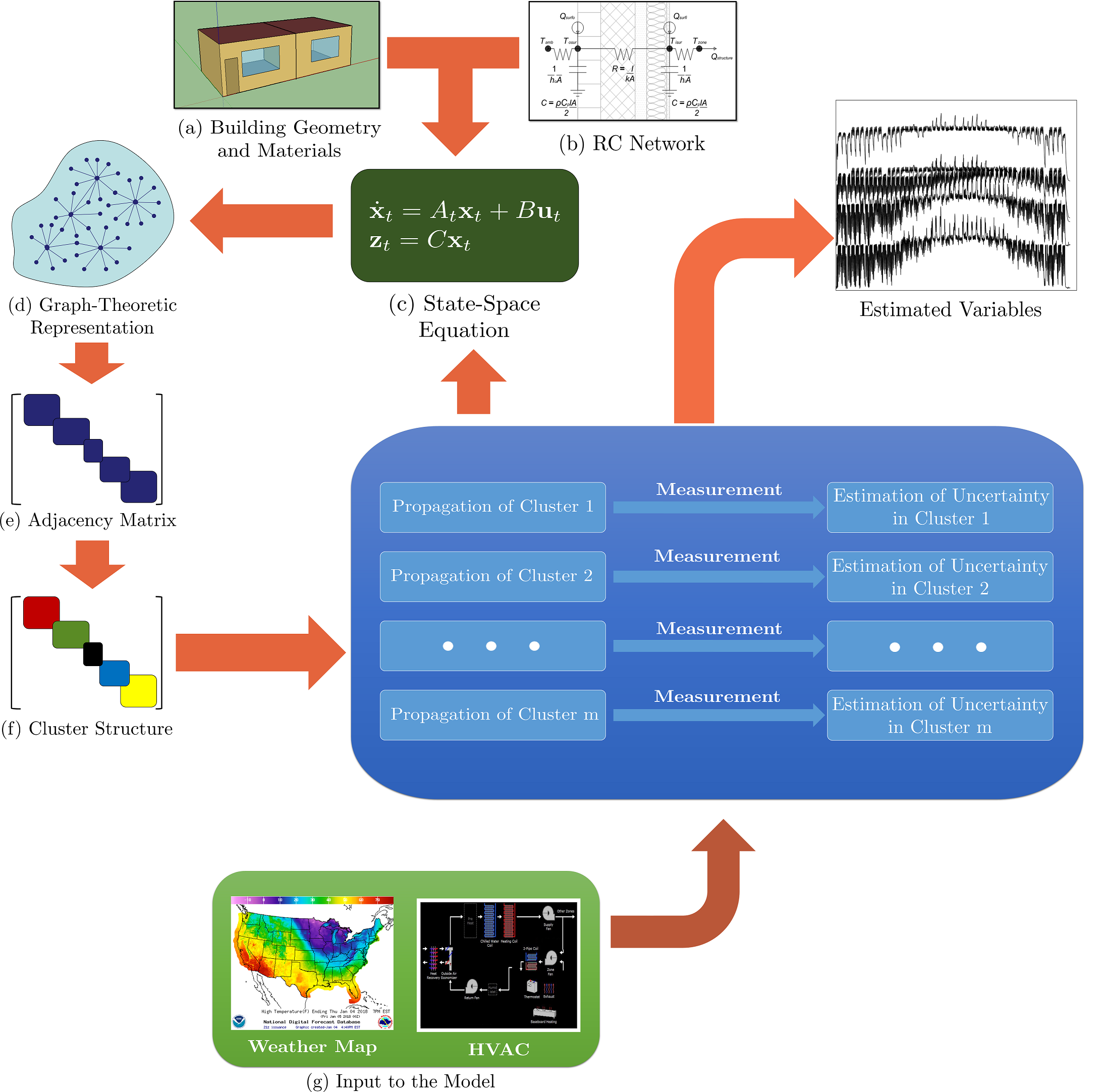}
\caption{Schematic of the WCS identification-based UQ to estimate thermal load in a large-scale BEM. Weather Map~\cite{weathermap}}
\label{fig:bem_framework}
\end{figure}

Figure~\ref{fig:bem_framework} depicts the overall UQ framework for a large-scale BEM. Given an office/school building (Figure~\ref{fig:bem_framework}(a)), the geometry and the thermal properties of the building are assessed to formulate the state-space equation model of a dynamical system (Figure~\ref{fig:bem_framework}(c)). This state-space equation is formulated using the concept of RC network (Figure~\ref{fig:bem_framework}(b)). The output equation is framed depending on the available measurement. A graph-theoretic representation is adopted to model the thermal network as an undirected graph (Figure~\ref{fig:bem_framework}(d)). The state-space equation and the initial uncertainty information is used to quantify the adjacency information for the undirected graph (Figure~\ref{fig:bem_framework}(e)). A suitable graph clustering algorithm (Louvain modularity optimization) is then implemented to identify the Weakly Connected Subsystems or WCSs (Figure~\ref{fig:bem_framework}(f)). Subsequently, the input formulation involving the weather (ambient and soil temperature) is obtained, and the solar gain for the exterior surfaces and HVAC airflow and temperatures are calculated, using eQUEST (Figure~\ref{fig:bem_framework}(g)). The statistical properties of the state variables are propagated through each WCSs, which are updated based on the measurement availability. The updated statistical properties give us the estimated parameters such as non-envelope load and the solar gain for internal surfaces, along with the zonal temperatures, and the surface temperatures. In the subsequent sections, the individual components of the overall framework are described in detail.

\subsection{Parameters of BEM: Geometry and Thermal Properties}
\label{Parameters}

The simulation requires the specification of the building geometry and surface properties.  A BEM is divided into thermal zones, based upon the actual HVAC system. Each area with an individual temperature sensor is considered as a thermal zone.  Within the zone, the following properties of each surface is determined:

\begin{enumerate}
\item Exterior Surface: Indicates surface is exposed to outside air conditions.
\item Interior Surface: Indicates surface is adjacent to another zone and exposed to the conditions of that zone.
\item Underground Surface: Indicates surface that is below the ground level and exposed to soil temperatures.
\item \textit{Adjacent Zone:} If interior surface, the identifier of the adjacent zone connected to the interior surface.
\item \textit{Area ($A^j$):} The net area of the surface, for windows $w = 1$ to $c$ on surface $j$ is derived as:
\begin{equation}
A^j = A_{gross}^j - \sum_{w=1}^c{A_{win}^{w}}
\end{equation}
\item \textit{Resistance ($R^j$):} A measurement of the heat flow across a surface at steady state conditions, calculated as follows:
\begin{equation}
R^j = \sum_{g=1}^h\frac{l^{g}}{k^{g}A^j}
\end{equation}
\noindent where $l$ is material thickness and $k$ is thermal conductivity for material $g = 1$ to $h$ in the construction of surface $j$.
\item \textit{Capacitance ($C^j$):} A measurement of the amount of heat input required to raise the temperature of a material. Capacitance encapsulates the capability of a material to store heat and delay heat transfer across the material.  $C^j$ is calculated as follows:
\begin{equation}
C^j = \sum_{g=1}^h\frac{\rho^{g}C_p^{g}l^{g}A^j}{2}
\end{equation}

\noindent where $l$ is material thickness, $\rho$ is density, and $C_p$ is specific heat for material $g = 1$ to $h$ in the construction of surface $j$.
This formula assumes half the capacitance is applied to each the inside and the outside faces of surface $j$.  
\item \textit{Window Resistance ($R_{win}^w$):} Similar to $R^j$.  Windows are modeled using only the specified resistance.  The thermal capacitance effects are negligible~\cite{kircher2015lumped}.
\item \textit{Inside Surface Conductance ($h_i^j$):} A measure of heat flow at the interior side of a surface $j$.  Assumed to be static based on surface location and position ~\cite{kircher2015lumped,american20132013}
\item \textit{Outside Surface Conductance ($h_o^j$):} Similar to $h_i$, but at the exterior side of a surface $j$.
\item \textit{Zone Air Mass ($M_k$):} The air mass of zone $k$ at design conditions.
\item \textit{Specific Heat of Air ($c_p$):} The specific heat of air at design conditions.
\end{enumerate}

To calculate input parameters, explained later, the following properties of the building are also assumed:
\begin{itemize}
\item \textit{Location:} The geographic coordinates of the overall building.
\item \textit{Orientation:} The cardinal direction and angle of the surfaces.
\item \textit{Shading devices:} The size and position of any permanent sun shades.
\end{itemize}

Surface properties are critically important for the BEM since the heat transfer occurring through the surfaces is a fundamental concept for these calculations.  Boundary conditions, in particular, determine the extent that the weather conditions impact the zone.  In the case of underground surfaces, there is no adjacent airspace, so there is no outside surface conduction $h_o$.  Instead, a fictitious conductance ($1/{R_{fic}}$) is used (explained later in Section~\ref{statespace}). 

The values of the material properties are determined by testing.  In a building, the properties are assumed based upon either specific data published from the manufacturer, or, if unknown, typical values of materials are compiled in subject references \cite{american20132013}.  In some cases, values of typical surface constructions as a whole are published.

\subsection{Heat Balance Method}
\label{heatbalmeth}

The heart of the BEM problem is the fundamental heat balance equation for zone $k$~\cite{doe2016energyplus}:

\begin{equation}
\label{heat_balance}
\resizebox{0.9\textwidth}{!}{$
\displaystyle C_k\frac{dT_k}{dt} = \sum\dot{T}^{int} + 
\sum_jh_jA_j(T_{amb} - T_k) + \sum_i\dot{m}_{mix} c_{pa}(T_{k,i} - T_k) + \dot{m}_{inf}c_{pa}(T_{\infty} - T_k) + \dot{m}c_{pa} (U_{sa} - T_k)
$}
\end{equation}
where:
\begin{itemize}
\item $C_k\frac{dT_j}{dt}$ = energy stored in the zone air
\item $\sum\dot{T}^{int}$ = sum of the convective non-envelope internal loads
\item $\sum_jh_jA_j(T_{amb} - T_z)$ = convective heat transfer from the zone surfaces $j$
\item $\sum_i\dot{m}_{mix} c_{pa}(T_{z,i} - T_k)$ = heat transfer due to inter-zone air mixing with zones $i$
\item $\dot{m}_{inf}c_{pa}(T_{\infty} - T_k)$ = heat transfer due to infiltration of outside air
\item $\dot{m}c_{pa} (U_{sa} - T_k)$ = heat transfer due to the output of the HVAC system into the zone
\end{itemize}
\noindent The definitions of the relevant variables are listed in the Nomenclature and discussed later.  Equation~\ref{heat_balance} includes a number of simplifying assumptions~\cite{american20132013,fisher1997convective}.  These assumptions are typical of most modeling techniques, and allow the use of a linear equation:
\begin{enumerate}
\item Air temperature in the zone is well-mixed. The variation of the zonal temperatures is zero.
\item The surface temperature and irradiation is uniform.
\item Surface radiation is diffuse.
\item One-dimensional heat conduction through the surface construction.
\end{enumerate}

\noindent Some other assumptions made for simplicity, which are typical in many BEMs, are as follows:
\begin{enumerate}
\item \textit{No inter-zone mixing.}   The supply and return airflows are balanced per zone, so the overall pressure differential of a building is approximately zero.  The majority of spaces are separated by doors.  Therefore, there is minimal air transfer between thermal zones, and this expression can be eliminated from the heat balance equation.
\item \textit{Simplified infiltration.} As discussed later, infiltration is not calculated directly and is included with other internal loads in the space as part of the lumped non-envelope load, eliminating this expression from the heat balance equation.
\item \textit{No long-wave radiation.} The air is assumed to be transparent.  Long-wave radiation exchanges between surfaces can be ignored  ~\cite{american20132013}.
\end{enumerate}

Latent loads, such as moisture due to the occupants, are not addressed in the BEM.  Humidity is not often directly controlled, but instead is handled as a side-effect of the sensible (dry-bulb) cooling.  Humidity primarily affects occupant comfort and the HVAC system efficiency, although a small amount of heat transfer occurs due to moisture in the air, with minimal impact on the zonal temperatures ~\cite{american20132013}.  Any simplification due to the model itself will be captured in the lumped non-envelope load estimation. Thus latent loads are ignored.

The reduced-order model as described in Section~\ref{statespace} is based upon the simplified heat equation as follows:

\begin{equation}
\label{simplified_heat_balance}
C_k\frac{dT_k}{dt} = \sum\dot{T}^{int} + \sum_jh_jA_j(T_{amb} - T_k) + \dot{m}c_{pa} (U_{sa} - T_k)
\end{equation}
  
\subsection{State Space Equation Using RC Network}
\label{statespace}

\begin{figure}[H]
\centering
\includegraphics[scale=0.6]{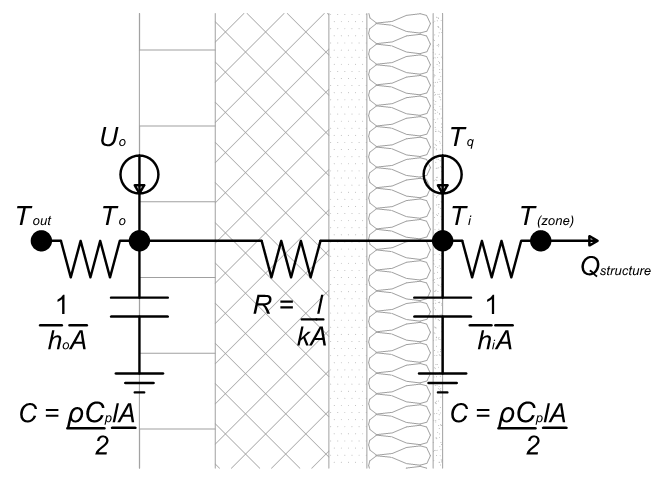}
\caption{RC network for surface constructions}
\label{rc_walls}
\end{figure}

As explained previously, the resistor-capacitor estimation method reduces the building surfaces into a discrete number of resistances and capacitances.  A three resistance, two capacitance model (3R2C) is the most used one.  This type of thermal load estimation model has been found to estimate the effects of the building surfaces on a thermal zone with sufficient accuracy~\cite{kircher2015lumped}.  Using this method, the heat balance equation can be decomposed into distinct parts that enable us to perform the load optimization.  The RC network for the surface constructions used for modeling purposes is shown in Figure~\ref{rc_walls}.

Summarizing the assumptions in Section~\ref{heatbalmeth} with the RC network model, the relevant heat balance equations are as follows for surface $j$ and window $w$ in zone $k$~\cite{o2010model}:
\begin{equation}
\begin{array}{c}
\label{surfeq}
\displaystyle C^j\frac{dT_o^j}{dt}=h_o^jA^j(T_{amb}-T_o^j)+\frac{T_i^j-T_o^j}{R^j}+U_o^j \\
\displaystyle C^j\frac{dT_i^j}{dt}=h_i^jA^j(T_k-T_i^j)+\frac{T_o^j-T_i^j}{R^j}+T^j_q \\
\displaystyle Q_{structure}=h_i^jA^j(T_i^j-T_k)+\frac{T_{amb}-T_k}{R_{win}^w}
\end{array}
\end{equation}

Underground zones are treated as special cases, as their exterior surfaces are not directly exposed to the outdoor temperatures. Instead, the majority of the heat transfer in the underground surfaces takes place at the exposed perimeter region of the surface. The constructions of the exterior surfaces of underground zones are modeled a no-mass R-value layer to avoid any overestimation of heat loss~\cite{winklemann2003underground}. The overall effective R-value is based on published data regarding the heat transfer through underground surfaces, taking into account the depth of the surface and the location and thickness of the insulation ~\cite{americanenergy}. 
\begin{equation}
R_{fic}^j = R_{eff}^j-R^j
\end{equation}
where $R_{eff}$ is the effective R-value, and $R$ is calculated as described in Section~\ref{Parameters}, including a 0.3 meter layer of soil as part of the surface construction. 

\begin{figure}[H]
\centering
\includegraphics[scale=0.6]{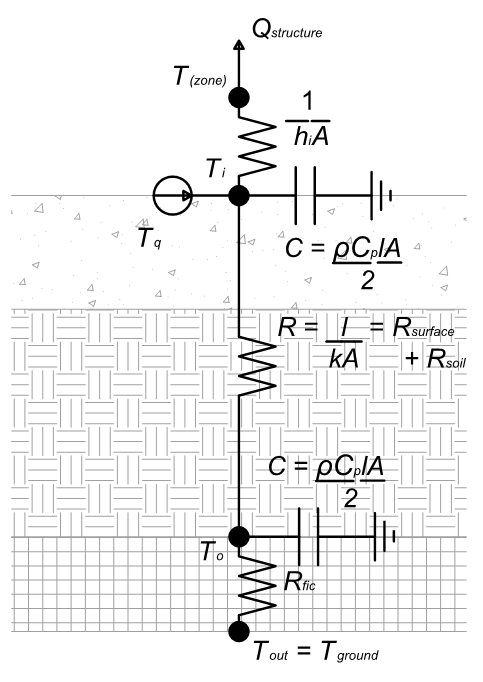}
\caption{Modified RC network for underground surfaces}
\label{rc_underground}
\end{figure}

The modified RC network for an underground surface is shown in Figure ~\ref{rc_underground}.

Using this method, one can use consistent equations for any surface type.  For the interior zones, the adjacent zone temperature $T_{out}$ is a calculated state variable. The $T_{out}$ for exterior surfaces is the ambient temperature $T_{amb}$ that is available from weather data. The ground or the soil temperature is used as the ambient temperature $T_{amb}$ of underground surfaces.

Combining Equations~\ref{surfeq}, the heat balance equation is modified as a system of Ordinary Differential Equations (ODE). The state space matrix derived from the ODE for the BEM can be written as:
\begin{equation}
\label{state_space}
\begin{array}{ll}
\dot{\textbf{x}}_t &= H_t \textbf{x}_t + B\textbf{u}_t \\
\textbf{z}_t &= C  \textbf{x}_t
\end{array}
\end{equation}

\noindent The state-space Equation~\ref{state_space} comprises of the zonal temperature $T_k$  and the inner and outer surface temperatures $T_i^m$ and $T_o^m$'s. In addition, for each zone, the solar gain $T_q^j$ for each inner surface and the non-envelope load $T_k^{ne}$ are modeled. Thus, $\textbf{x} \in \mathbb{R}^N$ can be decomposed into the collection of zonal variables as $\textbf{x} = \lbrace \textbf{T}_1, \textbf{T}_2, \ldots \textbf{T}_N \rbrace$. Each $\textbf{T}_k \in \mathbb{R}^{n_k}$, $k = 1$ to $N$ represents the collection the zonal temperature $T_k$, inner and outer surface temperatures for the $m_k$ surfaces, and the three load variables in a particular zone. Hence, $\textbf{T}_k$ can be written as:

\begin{equation}
\textbf{T}_k = \lbrace T_k, T_o^1, T_i^1, T_q^1, \ldots, T_o^{m_k}, T_i^{m_k}, T_q^{m_k}, T_k^{int}  \rbrace \hspace{5mm} n_k = 2 + 3m_k 
\end{equation}

\noindent And,

\begin{equation}
N = \sum_k^n n_k = \sum_k^n 3m_k + 2 = 2n + 3 \sum_k m_k
\end{equation}

\noindent $T_k^{ne}$ and  $T_q^j$ for each surface are modeled as state variables and are estimated with available measurement. Each zone temperature is determined by the conduction through the surface constructions, as well as the HVAC airflows and temperatures, surface solar gains, and internal non-envelope loads. Surfaces adjacent to another zone share properties with that zone. Zones with differing occupancies and internal loads are expected to have the most zonal interactions. The components of the zonal variable $\textbf{T}_k$ are modeled as~\cite{o2010model}:

\begin{equation}
\label{zone_equation}
\resizebox{\textwidth}{!}{$ 
\begin{array}{ll}
\dot{T}_k &= \displaystyle \left[ -\frac{\dot{m}_k}{M_k} - \frac{\sum_m A^m h_i^m}{M_k c_{pa}} -  \frac{ \sum_w \frac{1}{R_{win}^w}}{M_k c_{pa}} \right] T_k + \frac{\sum_m A^m h_i^m T_i^m}{M_k c_{pa}} + \frac{1}{M_k c_{pa}} T^{int}_k + \frac{\frac{1}{R_{win}^w}}{M_k  c_{pa}}  T_{amb} +\frac{\dot{m}_k U_k^{sa}}{M_k  c_{pa}}  \\
\dot{T}^j_o &= - \displaystyle \left[ \frac{h_o^j A_j}{C^j} + \frac{1}{R^j C^j} \right] T_o^j + \frac{1}{R^j C^j} T_i^j + \frac{h_o^j A_j}{C^j} T^j_{out} + \frac{U^j_o}{C^j}  \\
\dot{T}^j_i &= \displaystyle \frac{h_o^j A_j}{C^j} T^j_{k} + \frac{1}{R^j C^j} T_o^j  - \displaystyle \left[ \frac{h_i^j A_j}{C^j} + \frac{1}{R^j C^j} \right] T_i^j  + \frac{T^j_q}{C^j} \\
\dot{T}_k^{int} &= 0, \hspace{3mm}  \dot{T}^j_q = 0 \hspace{15mm} j = 1,\ldots,m_k
\end{array}
$}
\end{equation}

\noindent where:
\begin{itemize}
\item[] $\dot{m}_k$ = mass flow of supply air from HVAC system to zone $k$
\item[] $U_k^{sa}$ = temperature of supply air from HVAC system to zone $k$ 
\item[] $M_k$ = mass of air in zone $k$
\item[] $c_{pa}$ = specific heat of air
\item[] $T_k^{ne}$ = heat gain from non-envelope internal loads
\end{itemize}
The measurement to this system $\textbf{z} \in \mathbb{R}^n$ are the zonal temperatures $T_k$'s and is characterized by the observation matrix $C \in \mathbb{R}^{N \times n}$. 

\subsection{Input Formulation}

The majority of the inputs for Equation~\ref{zone_equation} can be obtained directly for zone $k$ at each time step.

For the outdoor conditions, Typical Meteorological Year (TMY) dataset is used for the nearest city~\cite{marion1995user}. The TMY data is not representative of any particular year but is developed to represent the typical conditions of a given location.  Estimated average monthly ground temperatures at different geographic locations are also available in the TMY dataset.  For a building with a Building Management System, the $T_{amb}$ conditions would be available with the HVAC trend data.

Solar heat gain on opaque exterior surface $U_o^j$, $j = 1$ to $m_k$ is calculated based on the buildings location and orientation in conjunction with the solar radiation intensity ~\cite{doe2016energyplus}:  

\begin{equation}
U_o^j = \alpha^j(I_b\frac{A_s^j}{A^j}\cos\theta^j + I_sF^j_{ss} + I_gF^j_{sg})
\end{equation}

The intensities of radiation are not measured directly but can be calculated from measured radiation (direct normal and diffuse horizontal) using the luminous efficacy models ~\cite{perez1990modeling}.  The Perez model is the basis of many modern solar models, including the approximate values given in the TMY dataset~\cite{marion1995user}. 
Since the real-time solar data is not available, the calculated input values for the surface solar heat gain are not dependent on actual weather conditions. Thus, for simplicity, the solar load values $U_o^j$ calculated by eQUEST are used in the model. Only exterior surfaces are subject to direct solar radiation and are assumed to have outer solar heat gain.

Calculating the HVAC system inputs for the heat balance equation requires specific information on fan curves, coil properties, pump curves, pressure losses, control strategies, capacity curves, etc.  Furthermore, there is a feedback loop relationship between the HVAC output and the building load components~\cite{he2016simplified}. Because of the complexity in calculating the air-side HVAC parameters and because the data required for this analysis are typically readily available with a robust building controls system, the hourly air temperatures and flows calculated by eQUEST are used in our state equations.

\subsection{Estimation of Solar Gain and Non-Envelope Loads}
In this work, the solar heat gains $U_o^j$ on surface $j$ and Non-Envelope Load $T^j_q$ are estimated through the model (as explained in Section~\ref{WCS}).  In a fully-functional BEM, such as simulated with eQUEST, both must be assigned approximate values to perform the simulation.

In such a BEM, the solar effect on the interior walls $U_o^j$ due to windows is calculated using the solar heat gain coefficient (SHGC).  First, the total heat gain through the windows $Q_{isol}^k$ for windows $w=1$ to $c$ in zone $k$ is calculated~\cite{american20132013}.
\begin{equation}
Q_{isol}^k=\sum_{w=1}^c(SHGC^w)A^wE_t^w
\end{equation}
\noindent where:
\begin{itemize}
\item[] $SHGC^w$ = Solar Heat Gain Coefficient of window $w$
\item[] $A^w$ = area of window $w$
\item[] $E_t$ = incident total irradiance of window $w$
\end{itemize}

The SHGC of the window represents the portion of solar radiation transmitted directly through the window, as well as the absorbed solar radiation.  This value is typically provided by the window manufacturer or can be assumed to be based on the window properties~\cite{american20132013}. From $Q_{isol}^k$, the solar flux $T_q^j$ on the interior surface $j$ can be estimated, using one of several approaches.  One such approach, 
is to assume that all radiation first hits the floor of zone $k$, and is reflected evenly across all the surfaces $r=1$ to $s$ in zone $k$~\cite{o2010model}.  In the current work, this internal solar gain $T_q$ is estimated as a part of the UQ framework.

Non-envelope internal loads $T_k^{ne}$, which refer to loads in the space not related to the building exterior walls, are also unknown and are estimated through the WCSs-based UQ framework. in a full BEM, load schedules are assumed, such as used by He at al.~\cite{he2016simplified}.  It is unlikely, however, that any physical building would follow the exact prescribed schedules with precision. Thus an optimal estimation method becomes necessary, especially due to the involved high-dimensional system.

In a full simulation, the airflow due to infiltration must be assumed.  The infiltration rate at any given time is based upon the pressure differential between the building and the exterior environment, as well as the effective leakage area of the building.  
%
These parameters depend on air temperature, air density, wind speed, and wind direction, and the resulting differential is modified by the nature of the building openings.  To obtain an accurate, effective leakage area, a blower door test is required, pressurizing the space to a specified value ~\cite{american20132013}. For a large building, this is a costly process.  Consequently, the estimated peak infiltration rate is usually simply based on an assumed construction tightness.  Because of the difficulty obtaining accurate infiltration data, the effect due to infiltration is also estimated with the other internal loads as part of the non-envelope load estimation.

Most thermal zones also have furniture that provides thermal mass and additional surfaces for radiation.  Specific information on the furniture in the zone, especially thermal properties, is difficult to obtain.  Wang and Xu~\cite{wang2006parameter} use a 2C2R model to estimate the internal mass of a zone using a genetic algorithm; however, the resulting parameters have no physical meaning other than an assumed lumped mass for estimation purposes.  Therefore, the state space equations in this work do not directly include any effects due to internal mass. Internal mass effects are implicitly included in the non-envelope load estimation. 

In the presented case study, there are zone-level HVAC units such as hot water baseboard radiation. These baseboard units provide additional radiant or convective heat to the zone.  The Building Management System (BMS) controls and monitors these baseboard units that work in conjunction with the supply air.  The baseboard units are included as a special case of the internal load. Like the HVAC parameters, the eQUEST hourly heat output values for the baseboard has been used.  In the case study, the BMS tracks the operation of the control valve, which can be used with the design flow rates and actual hot water loop temperatures to approximate the unit heat output.  This value is then simply subtracted from the calculated internal load.  

\subsection{Identification of Weakly Connected Subsystems and Optimal Estimation}
\label{WCS}

Optimal estimation of the unknown variables $T_i,T_o, T^{int}$ and $T_q$ in the problem detailed in Equation~\ref{zone_equation} refers to solving the following minimization problem
\begin{equation}
\displaystyle \min_{\hat{\textbf{x}}_k} E\left[|| \textbf{x}_k - \hat{\textbf{x}}_k ||_2 \right]
\end{equation}
\noindent The above term refers to the error in the \textit{a posteriori} state estimation. For a given measurement $\textbf{z}_t$, solution to the problem is same as solving the well known \textit{Filtering} problem. Consider the solution at time $t-1$ as $\hat{\textbf{x}}_{t-1}$ and covariance $\Sigma_{t-1}$, the estimate $\hat{\textbf{x}}_t$ and covariance $\Sigma_t$ are given as,

\begin{equation}
\label{kalman_full}
\begin{array}{l}
\hat{\textbf{x}}_t = \hat{\textbf{x}}_{t|t-1} + K\textbf{y}_{t} \\
\Sigma_t = \Sigma_{t|t-1} - K (R + C \Sigma_{t|t-1} C^T)  K^T
\end{array}
\end{equation}

\noindent where $K$ is known as the Kalman gain and $\textbf{y}_t = \textbf{z}_t - \hat{\textbf{x}}_{t|t-1}$ is known as the measurement residual. The \textit{a priori} estimates $\hat{\textbf{x}}_{t|t-1}$ and $\Sigma_{t|t-1}$ are one step solution to the Equation~\ref{state_space} depending on $\hat{\textbf{x}}_{t-1}$ and covariance $\Sigma_{t-1}$. The expression for the Kalman gain $K$ is given as~\cite{kalman1960new},

\begin{equation}
K = \Sigma_{t|t-1} C \left( R + C \Sigma_{t|t-1} C^T \right) ^{-1}
\end{equation}

Due to the high dimensionality of the problem involving a large number of state variables $N$, the series of matrix operations becomes computationally expensive. The above filtering problem is solved through Identification of Weakly Connected Subsystems (WCSs)~\cite{Mukherjee_2017,mukherjee2015laplacian,mukherjee2015non} to increase the computational efficiency. The WCS-based method is effective in solving high-dimensional UQ problems involving linear and non-linear filtering problems. 

WCSs for the state variable $\textbf{x} \in \mathbb{R}^N$ are the countable, mutually exclusive and exhaustive partitions $\textbf{y}_j  \in \mathbb{R}^{n_j}$, such that the following relation holds 
\begin{equation}
\begin{array}{l}
\displaystyle P(\textbf{x}_t) = \prod_j P(\textbf{y}_{j_t}) \\ 
\sum_j n_j = N
\end{array}
\end{equation}

The index $t$ represents that the relation is invariant under the transformation given by Equation~\ref{state_space} for a time-period $t \in [0, T)$. Performing such decomposition of $\textbf{x}_t$ enables faster UQ by solving parallel subproblems given as:

\begin{equation}
\label{sub-problem}
\dot{\textbf{y}}_{j} = H_j(t)\textbf{y}_{j} + B_j \textbf{u}_j
\end{equation}

The solutions to each subproblem in Equation~\ref{sub-problem} is given by the following continuous time Kalman filter~\cite{jazwinski2007stochastic}:

\begin{equation}
\begin{array}{l}
\displaystyle \dot{E(\textbf{y}_{j})} = H_j(t)E(\textbf{y}_{j}) + B_j \textbf{u}_j + K_j(z_j - C_j E(\textbf{y}_{j_t})) \\
\displaystyle \dot{\Sigma_j} = H_j(t)\Sigma_j + \Sigma_j H_j(t)^T - K_j R_j K_j^T \\
\displaystyle K_j = \Sigma_j H_j^T R_j^{-1}
\end{array}
\end{equation}

\noindent This invokes the use parallel computation for both the one-step \textit{a priori} estimation and as well the use of Kalman filter for the \textit{a posteriori} estimation. The state estimates $\textbf{x}_t$ and $\Sigma_t$ are computed from the WCSs using \textit{direct sum} of the vector spaces as:

\begin{equation}
\label{jbs:moment_cl}
\begin{array}{l}
E(\textbf{x}_t) = E(\textbf{y}_{1_t}) \oplus E(\textbf{y}_{2_t}) \oplus \ldots \oplus E(\textbf{y}_{m_t}) \\
\Sigma_t = \displaystyle \bigoplus_j \Sigma_{j_t} = \text{diag}\left( \Sigma_{1_t}, \ldots, \Sigma_{m_t}  \right)
\end{array}
\end{equation}

The normalized symmetrized adjacency matrix derived from the state-space matrix $H = \left( h_{ij}  \right)$ in Equation~\ref{state_space}~\cite{Mukherjee_2017} is given as:

\begin{equation}
\label{symm}
W = 0.5 (D^{-1} A_{abs} + A_{abs}^T D^{-1})
\end{equation}

\noindent where, $A_{abs} = \left( | h_{ij} | \right)$ and $D$ is the corresponding degree matrix of $A_{abs}$. The clusters are identified using Louvain method of community detection~\cite{blondel2008fast}. The method identifies weakly connected components in a weighted graph by maximizing modularity function defined as:

\begin{equation}
Q = \frac{1}{2m_q} \displaystyle \sum_{i,j} \left[W_{i,j} - \frac{k_i k_j}{2m_q} \right] \delta(c_i, c_j)
\end{equation}

\noindent where, $m_q = \sum_{i,j}W_{i,j} = N/2$, $k_i = \sum_i W_{i,j}$ and $c_i$ is the partition to which $i^{\text{th}}$ state belongs. The delta function $\delta(c_i,c_j)$ is 
\begin{equation}
\delta(c_i,c_j) = \begin{cases}
1 & c_i = c_j \\ 0 & \text{otherwise}
\end{cases}
\end{equation}
Maximizing $Q$ gives the values of $c_i$'s, $i = 1$ to $N$ and hence determines the cluster structure. In the next section, details pertaining to a building used for BEM modeling purposes are outlined.

\section{BEM Details}
\label{case_study}

\subsection{Building Description}
\label{description}

\begin{figure}[H]
\centering
\includegraphics[scale=0.4]{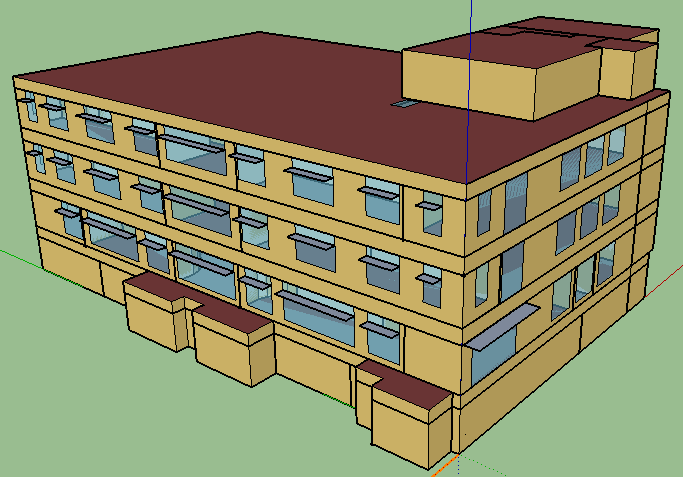}
\caption{Schematic of College/University building in Central New York used for creating the case study BEM}
\label{building}
\end{figure}

The case study building used in this work to illustrate the efficacy of the outlined UQ framework is a College/University building in Central New York, United States (see Figure~\ref{building}). The facility is an existing 4-story 5,050 square meter building with a mechanical penthouse.  It is comprised of primarily classrooms and offices, student lounges, conference rooms, observation rooms and as well as other support spaces.  The lowest floor comprises of underground zones, and the northeast portion of the building is attached to an adjacent structure.

The design of the building HVAC system informs the eQUEST calculations for the inputs for $U^{sa}$ and $\dot{m}$.  The individual spaces are conditioned at the zone level by fan-coil units with fin-tube radiation on the perimeter.  A dedicated outdoor air system supplies tempered ventilation air directly to the fan-coil units.  The hot and chilled water coils are supplied from the campus plants.  Network data rooms are conditioned separately with a variable refrigerant flow heat pump system.

Surface constructions determine the model parameters $R$ and $C$.  The brick and concrete block envelope has a combination of rigid and spray-applied insulation at the exterior walls, and the reflective membrane roof is concrete deck topped with rigid insulation. The windows are tinted high-performance glazing with sunshades on the southern exposure.

To understand $T^{int}$, the lighting, occupancy, and equipment must be considered.  The building uses high-efficiency LED lighting with occupancy sensors and daylighting controls. The building operates Monday through Friday from 7 am to 10 pm, with the expected use of a typical university building.

\begin{figure}[H]
\centering
\includegraphics[scale=0.4]{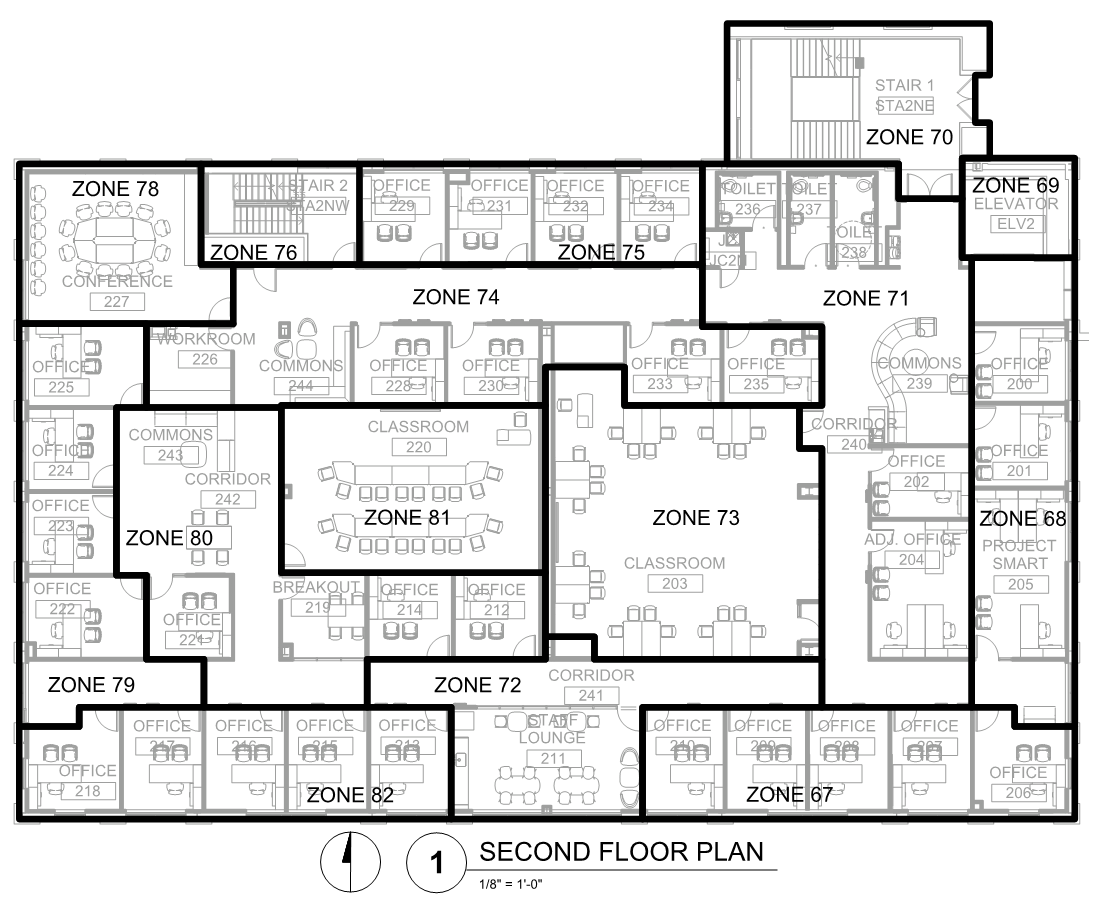}
\caption{Second Floor Thermal Zones of the BEM}
\label{floor_plan}
\end{figure}

There are a total of 132 thermal zones in the building, including unconditioned plenum spaces above the ceilings, for 61 directly conditioned zones.  In the modeled building, there are 668 interior surfaces, 69 underground surfaces (including slab-on-grade floors), 124 exterior surfaces (including roofs), and 80 windows.  The second-floor thermal zones are shown in Figure~\ref{floor_plan}.

Four zones are specifically considered to illustrate the details of our methodology. These are a ground floor perimeter zone with underground walls (Zone 1), a first floor exterior zone with west facing windows (Zone 47), a second floor zone with no exterior surfaces (Zone 80), and a third floor zone with a roof and south and east exterior walls (Zone 97).  The number of input parameters needed per zone to calculate only conduction through building surfaces is listed in Table~\ref{zonetable}.  Each zone also has dynamic parameters $T_{int}^k$ for lighting, occupancy, equipment load, and infiltration, as well as the inputs $\dot{m}_k$ and $U_{sa}^k$ for the HVAC system and assumptions by the BEM program.  Each of these inputs has uncertainty associated with them. As these zones represent only 4 out of the 132 zones, it is clear that this is a high-dimensional problem that requires a scalable UQ method. 

\begin{table}[H]
\begin{center}
\caption{BEM Inputs per Zone - Conduction through Surfaces Only}
\resizebox{\textwidth}{!}{ 
\begin{tabular}{cccccccc}
\hline
& Number of & Number of & Number of & & Number of  & Number of & Number of \\ 
& Exterior & Interior & Underground & Number of & Construction & Adjacent & BEM Input \\ 
& Surfaces & Surfaces & Surfaces & Windows & Materials & Zones & Parameters\\ 
\hline
Zone 1 & 0 & 6 & 4 & 0 & 11 & 2 & 52 \\
Zone 47 & 2 & 12 & 0 & 2 & 11 & 3 & 80 \\
Zone 80 & 0 & 6 & 0 & 0 & 10 & 4 & 48 \\
Zone 97 & 2 & 8 & 0 & 6 & 14 & 4 & 81 \\
\hline
\end{tabular}
}
\label{zonetable}
\end{center}
\end{table}

The main floors of the modeled building are partially attached to an adjacent building.  However, the adjacent building is controlled independently of the modeled building, and the properties of the adjacent building are unknown. Therefore, the surface boundary conditions with the adjacent building are assumed to be adiabatic, with no heat transfer of any kind between the two buildings.  

The modeled building is used to demonstrate our proposed methodology largely due to its size and complexity.  With 61 directly conditioned thermal zones, one can explore the differences between and interconnectivity of perimeter zones and core zones.  There are several exterior wall construction types, as the first floor differs from the upper floors, and the underground walls require a modified approach to calculate the heat transfer.  Complex lighting controls create a further challenge, which compels the internal load estimation and UQ.  There are a large number of input parameters that necessitates the use of a scalable UQ method.


\subsection{Full-Scale BEM Details}
\label{equest}

\begin{figure}[H]
\centering
\includegraphics[scale=0.45]{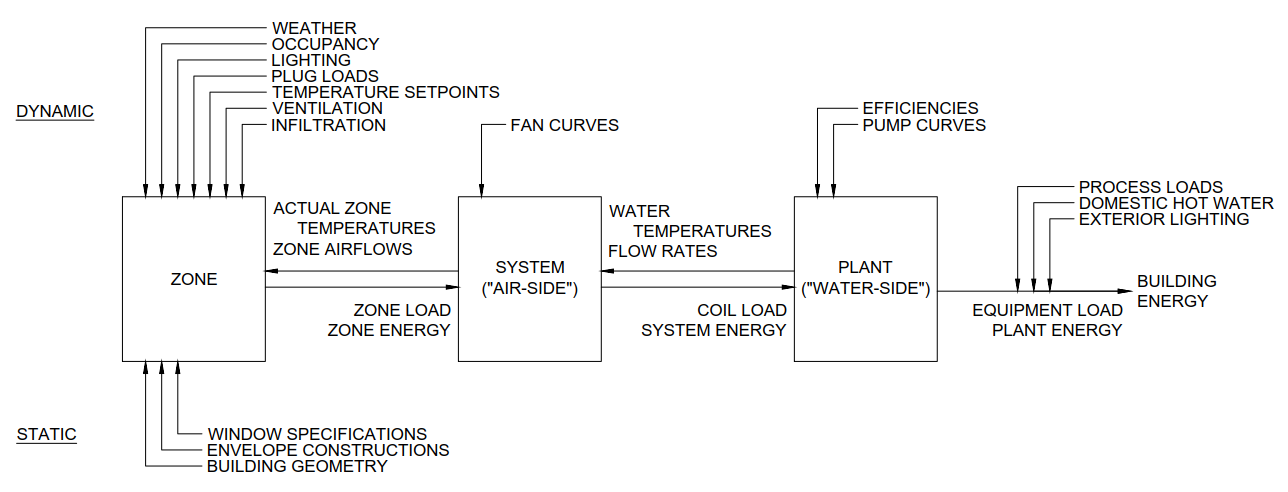}
\caption{Schematic for calculation of building energy}
\label{process}
\end{figure}

To provide data to our reduced-order model, this building has also been modeled in eQUEST (version 3.65)~\cite{hirsch2006equest} following the methodology laid out in ASHRAE 90.1-2013 (Appendix G ~\cite{american90}).  All known parameters of the building, as discussed in Section~\ref{case_study}, is provided as an input to the software program. eQUEST uses a white-box approach to simulate building energy. At first, it calculates the zone load at each time step. This is followed by converting the zone load into the required load and conditions of the HVAC system. The calculated load is fed back into the load calculation~\cite{hirsch2003doe}.  The process for calculating building energy use is depicted in Figure~\ref{process}. 

Since the modeled building has a digital building management system, one has access to the HVAC airflows and temperatures, water flow rates and temperatures, zone temperatures, as well as a variety of other control points. A list of actual measurements available for the fan-coil units in this building from the BMS is listed in Table~\ref{controls}. The input to Equation~\ref{state_space} is obtained directly from the simulation of the eQUEST model.

\begin{table}[H]
\begin{center}
\caption{Fan Coil Unit Control Points}
\small
\begin{tabular}{|l|c|c|c|c|}
\hline
& Analog & Analog & Digital & Digital \\ 
Description & Input & Output & Input & Output\\ 
\hline \hline
Fan Enable/Disable & & & & X\\
\hline
Fan Status & & & X & \\
\hline
Space Temperature & X & & &\\
\hline
Heating Valve Position & & X & & \\
\hline
Cooling Valve Position & & X & & \\
\hline
Leaving Air Temperature & X & & & \\
\hline
Filter Pressure Differential Sensor & & & X & \\
\hline
Condensate Alarm & & & X & \\
\hline
\end{tabular}
\label{controls}
\end{center}
\end{table}

In the eQUEST model, internal loads are assumed using hourly schedules. Lighting peak loads are calculated directly from the known lighting layout. However, the occupancy and receptacle loads are estimated.  Building code provides an approximation for the maximum occupancy for each space type~\cite{council2015international}.  Receptacle loads can be highly variable, especially with today’s proliferation of electronics. Typical representative load values have been compiled in COMNET~\cite{resnet2010}.  These peak load conditions are then scaled hourly based on estimated schedules.  The ASHRAE 90.1 User’s Manual provides typical fractional schedules for occupancy, zone lighting, and receptacle loads~\cite{americanenergy} that have been modified in this work. The building lighting in the modeled building is assumed to be controlled by a combination of daylight harvesting controls, occupancy sensors, dimming switches, and manual on/off controls.  Using simple reduction factors, ASHRAE 90.1 provides guidance on the expected effect of some of the lighting controls ~\cite{american90}.  However, outcomes are not representative of a particular building, and it is likely that in practice, the actual hourly lighting power density would be substantially different than the predicted.

\section{Results}
\label{results}
\subsection{Adjacency and Cluster Analysis}
\label{adjacency}

The BEM described in Section~\ref{case_study} contains 2649 state variables (Involving $\lbrace T_k,T_o,T_i,T_q,T^{int} \rbrace$), with 1125 input variables. The state-space equation (in Equation~\ref{state_space}) of the BEM is based on the thermal properties of the materials of the building and the interaction of the zonal and surface temperatures. The adjacency matrix $W$ is a reflection of coupling along with the presence of any inter-zone coupling. The variable $T_{out}^j$ in Equation~\ref{zone_equation} can be both the ambient temperature $T_{amb}$ or the temperature of adjacent zone. Thus, the inter-zone coupling (if present) depends on the magnitude of $h^j_o A_j/C_j$. Additionally, the adjacency matrix is highly sparse. This makes a visual representation of the whole $W$ very difficult to interpret. Results aggregated from the four thermal zones listed in Table~\ref{zonetable} representing diverse zonal conditions: Zone 1, Zone 47, Zone 80 and Zone 97 are discussed next.  

\begin{figure}[h]
\centering
\subfigure[Zone 1]{\includegraphics[width=0.4\textwidth]{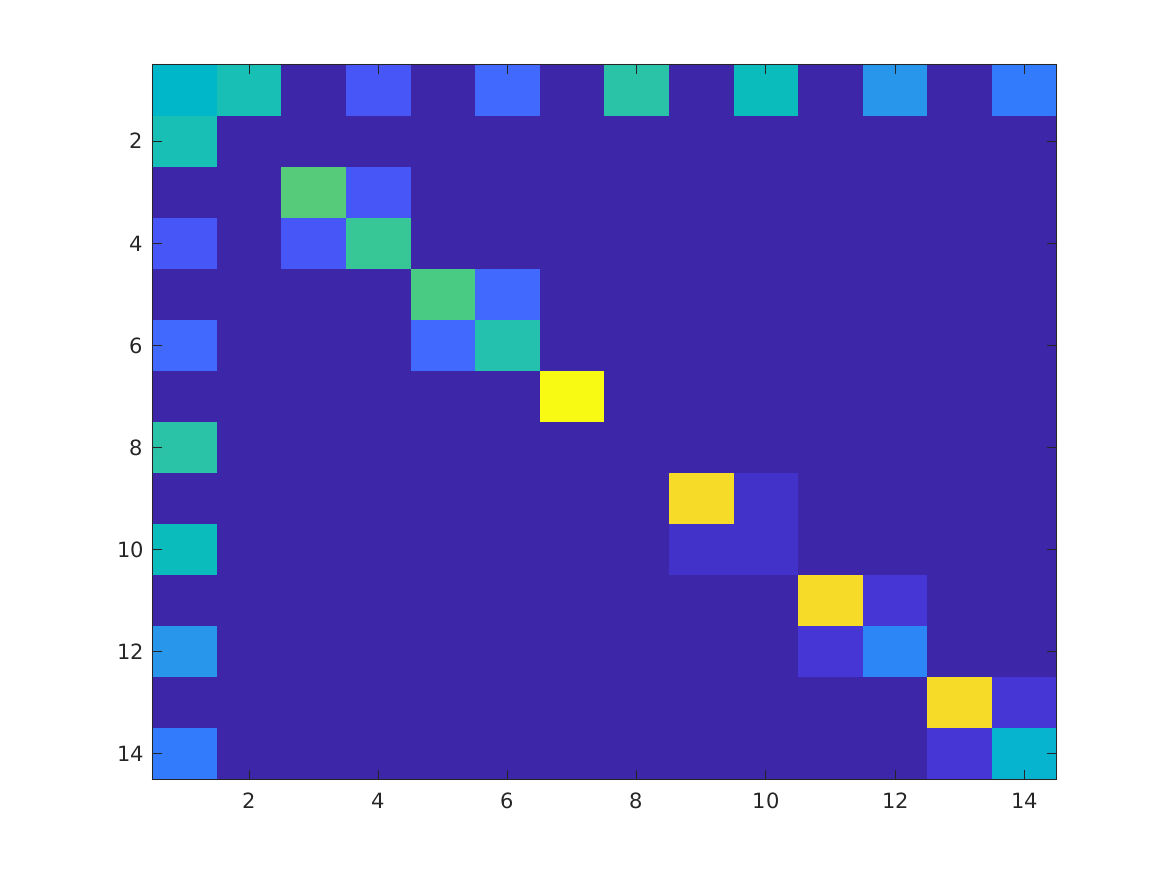}}
\subfigure[Zone 47]{\includegraphics[width=0.4\textwidth]{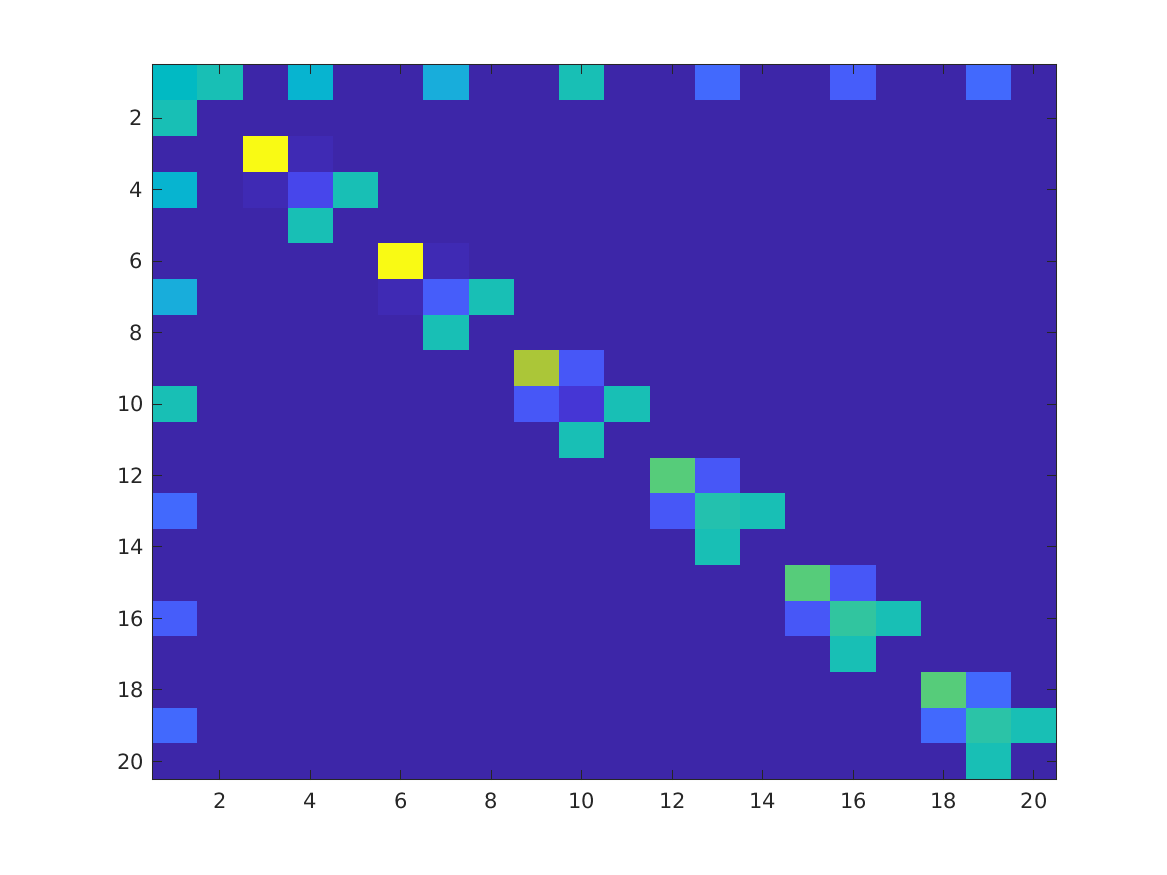}} \\
\subfigure[Zone 80]{\includegraphics[width=0.4\textwidth]{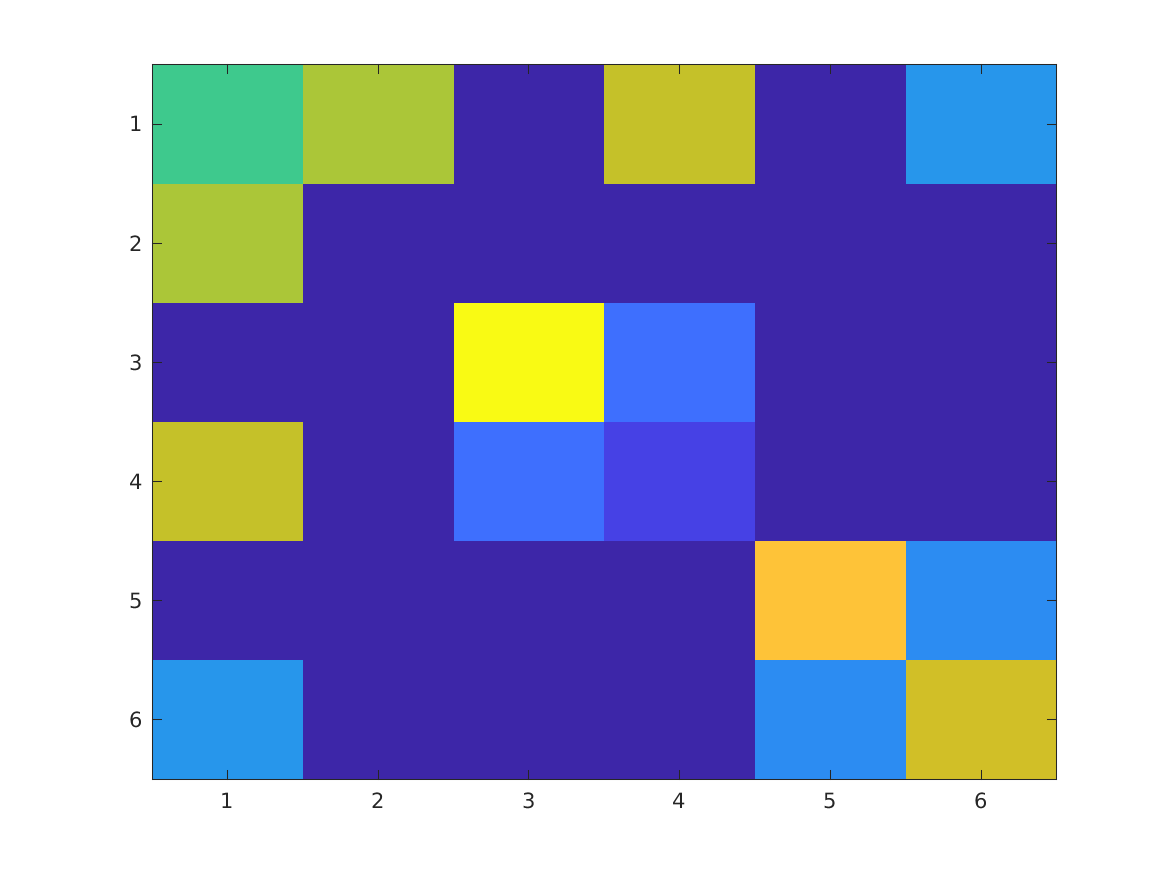}}
\subfigure[Zone 97]{\includegraphics[width=0.4\textwidth]{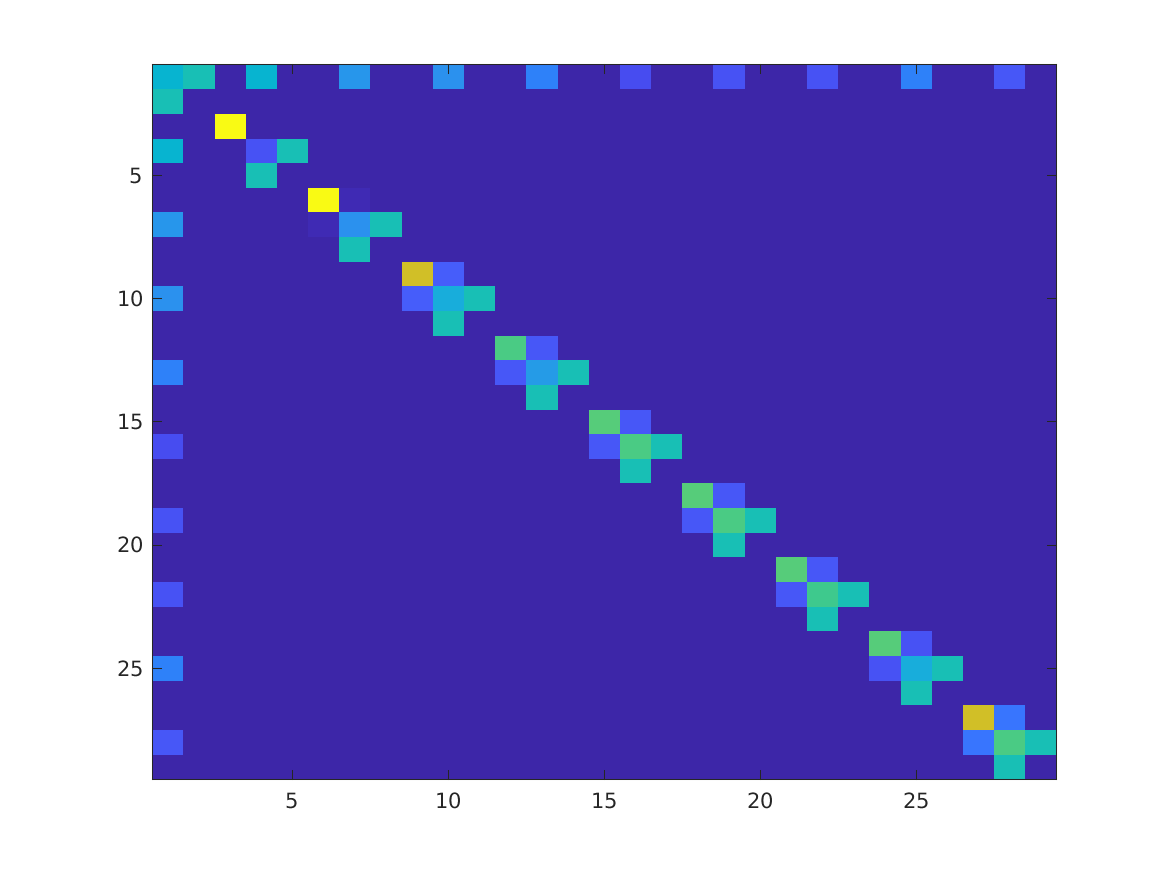}}
\caption{Adjacency Information for (a) Zone 1, (b) Zone 47, (c) Zone 80 and (d) Zone 97}
\label{fig:Zone_adjacency}
\end{figure}

\begin{figure}[H]
\centering
\includegraphics[width=0.8\textwidth]{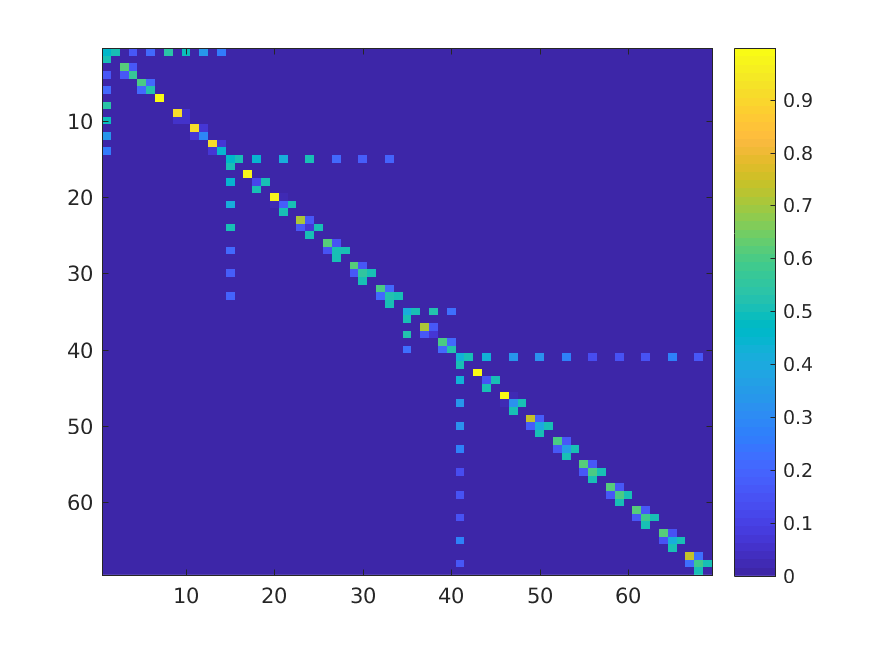}
\caption{Adjacency Matrix for Zone 1, Zone 47, Zone 80 and Zone 97 combined }
\label{adj_new}
\end{figure}

Figure~\ref{fig:Zone_adjacency} shows the sub-matrices of the Normalized Adjacency matrix $W$ corresponding to the four zones. The zones show the involved variables $T_k,T_o,T_i,T^{int}$. Zone 1 and 80 do not have exterior surfaces. Thus the variable $T_q$ is absent from the corresponding adjacency figures (Figures~\ref{fig:Zone_adjacency}(a) and (c)). Figure~\ref{adj_new} demonstrates the inter-zone coupling between these four zones. The plot shows no interaction between these four zones. However, this result does not conclude that there is zero interaction between other zones.

\begin{figure}[H]
\centering
\includegraphics[width=0.8\textwidth]{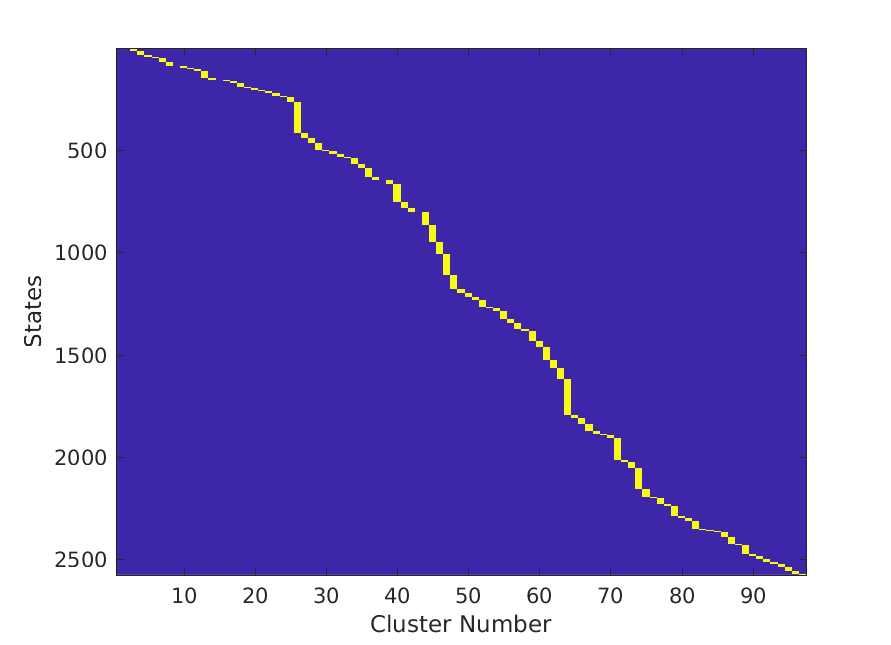}
\caption{Cluster Matrix}
\label{cluster_matrix}
\end{figure}

The cluster matrix is depicted in Figure~\ref{cluster_matrix}. 

The application of the Louvain modularity optimization-based clustering algorithm on the derived $W$ matrix for 2649 states resulted in identification of $m=97$. The number of identified clusters ($m=97$) is less than the number of zones(=132). However, the cluster structure works in accordance with the adjacency information identified in Figure~\ref{fig:Zone_adjacency} and~\ref{adj_new}. It can be observed in Figure~\ref{cluster_matrix} that consecutive states participate in clusters. Intuitively it can be concluded that the clustering algorithm either recognizes a single zone or a group of zones as a cluster. In most cases it ignores the inter-zone coupling. In the next section, the whole system is analyzed along with the estimation of the thermal loads via these identified clusters or WCSs. To compare the performance of the WCS-based UQ, the whole BEM is also analyzed without any clustering.

\subsection{Estimated Temperatures and Thermal Loads}

\begin{figure}[H]
\centering
\includegraphics[width=0.5\textwidth]{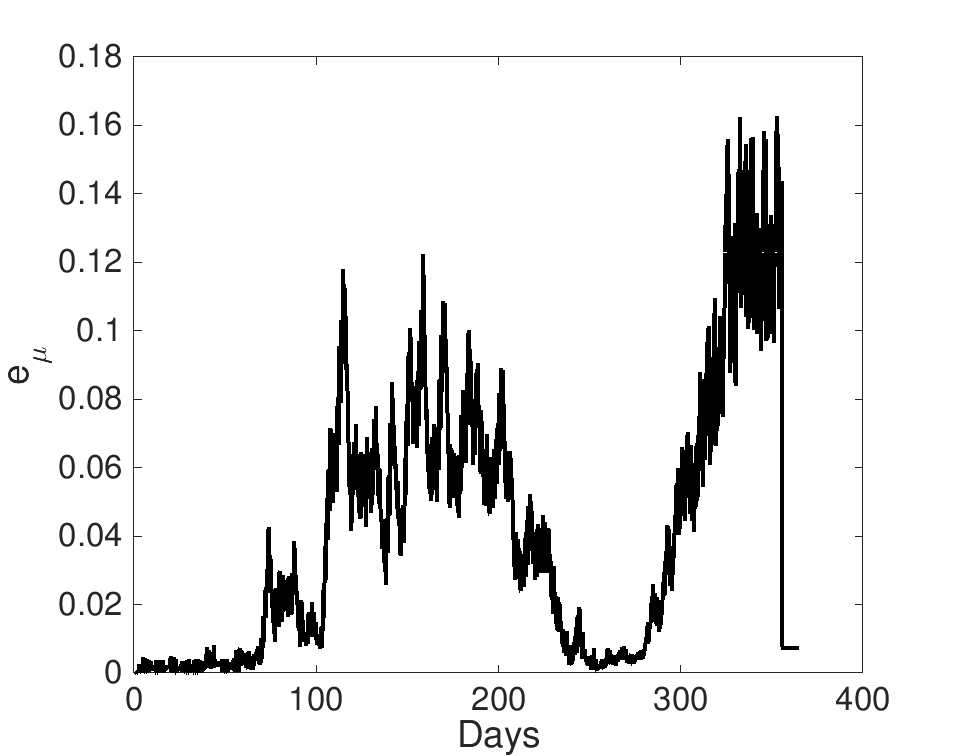}
\caption{Plot of error in estimation vs time}
\label{jbs:fig:error}
\end{figure}

For the simulation, a measurement noise is assumed for each of the zonal temperature as $R = \text{diag}\lbrace r_1, r_2, \ldots, r_{132} \rbrace$ where $\lbrace r_1, r_2, \ldots, r_{132} \rbrace$ are randomly generated numbers between 0 and $0.5$. This corresponds to the accuracy of typical room temperature sensor~\cite{siemens2013catalog}. 
Before displaying the estimated thermal loads, the accuracy of the WCS-based UQ algorithm is demonstrated first by a plot of error in estimation metric vs time. This metric is computed at each time as:

\begin{equation}
\label{jbs:err_metric}
e_{\mu} = \frac{1}{n}\begin{Vmatrix}
\frac{E(\textbf{x}_{t_c}) - E(\textbf{x}_{t_f})}{E(\textbf{x}_{t_f})}
\end{Vmatrix}_2
\end{equation}

\noindent where, $E(\textbf{x}_{t_c})$ is the mean of the state variable $\textbf{x}_t \in \mathbb{R}^n$ obtained from the Equation~\ref{jbs:moment_cl}. $E(\textbf{x}_{t_f})$ is obtained by running the full model (Equation~\ref{kalman_full}). Figure~\ref{jbs:fig:error} shows the plot of $e_{\mu}$ vs time for a span of one year.

\begin{figure}[H]
\centering
\subfigure[Zone 1]{\includegraphics[width=0.45\textwidth]{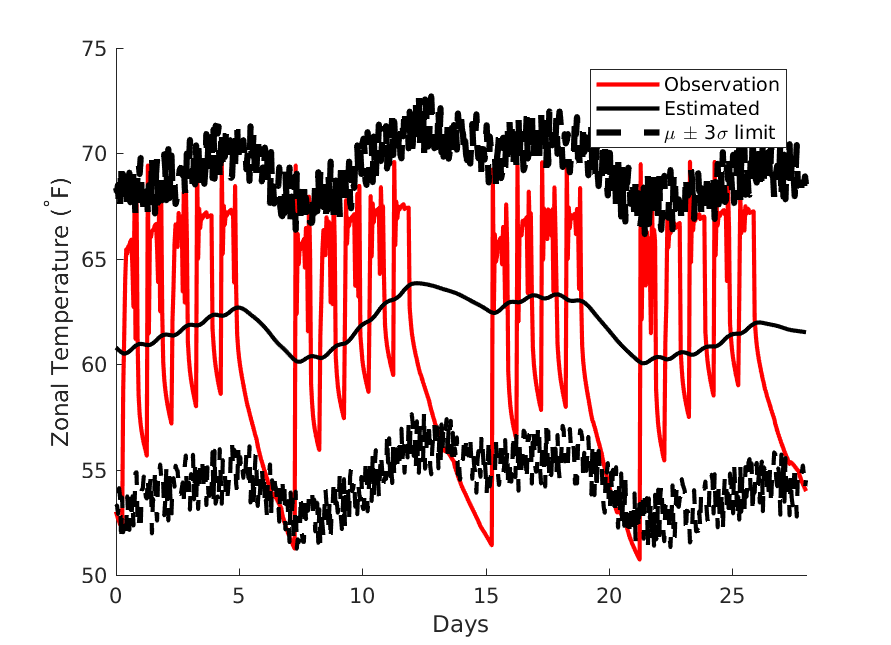}}
\subfigure[Zone 47]{\includegraphics[width=0.45\textwidth]{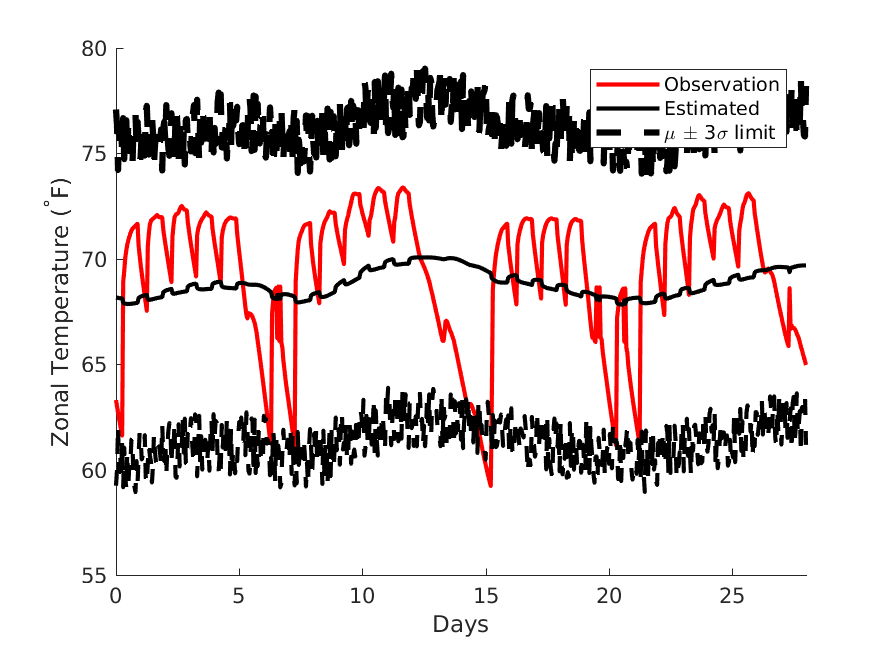}} \\
\subfigure[Zone 80]{\includegraphics[width=0.45\textwidth]{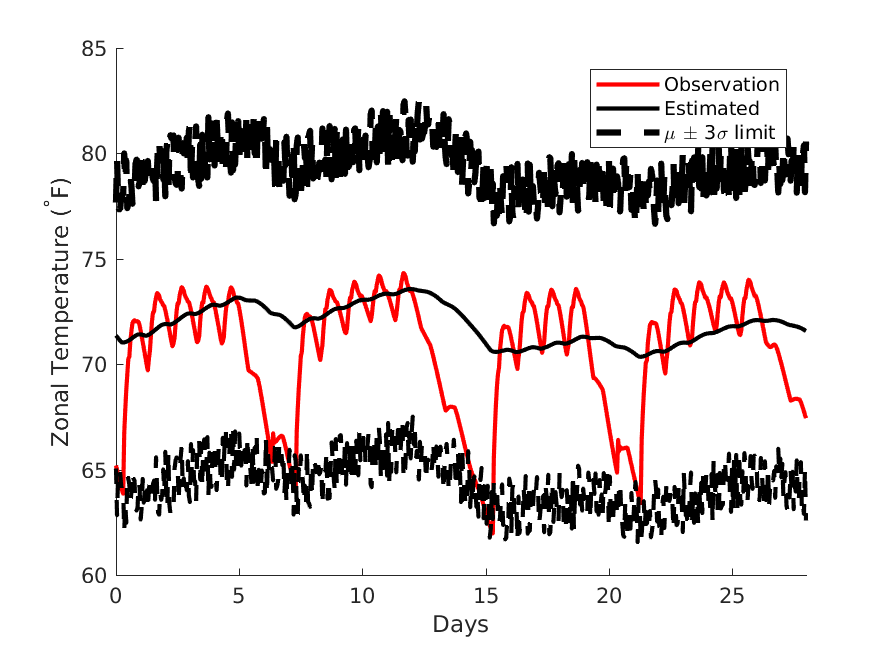}}
\subfigure[Zone 97]{\includegraphics[width=0.45\textwidth]{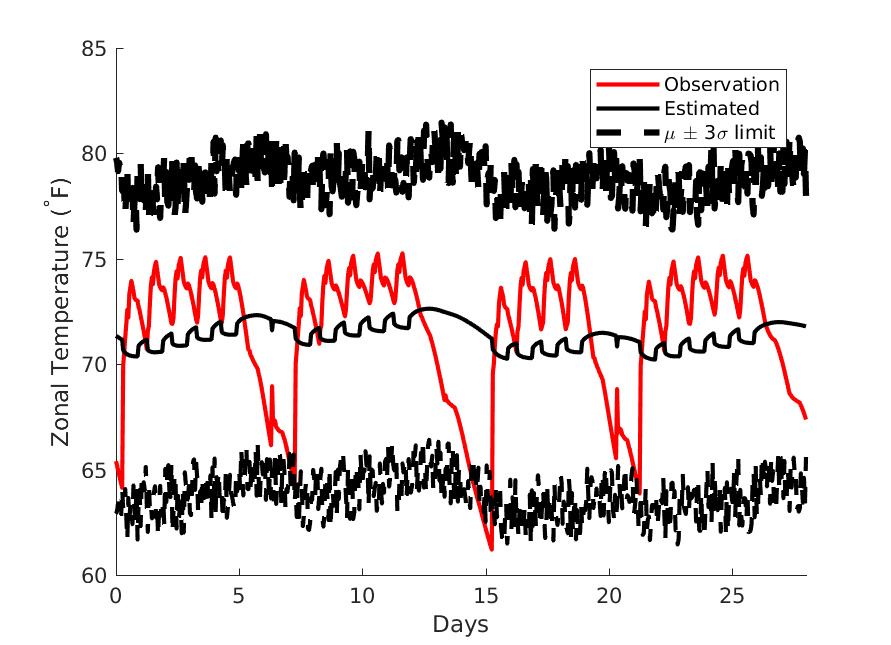}}
\caption{Adjacency Information for (a) Zone 1, (b) Zone 47, (c) Zone 80 and (d) Zone 97}
\label{Zone_temperature}
\end{figure}

Figure~\ref{jbs:fig:error} shows the accuracy of the WCS-based UQ approach. 
The identified WCSs ignore the inter-zone mixing of the surface temperatures. This mixing has very little effect on the results. The error metric $e_{\mu}$ converges to a value between 0.02 and 0.20 (2-20$\%$) towards the later part of the year. The slow convergence might be due to the presence of large number of state variables compared to fewer available observations. Also, the individual zonal temperatures are not an explicit function of the surface temperatures. This causes difficulty in the estimation of the surface temperatures $T_o$ and $T_i$ from $T_k$. Due to this convergence rate, subsequent analysis has been shown skipping the first month of the year. 

Figure~\ref{Zone_temperature} shows the accuracy in estimating the zonal temperatures for the four zones; Zone 1, Zone 47, Zone 80 and Zone 97.

\begin{figure}[H]
\centering
\subfigure[Zone 1]{\includegraphics[width=0.45\textwidth]{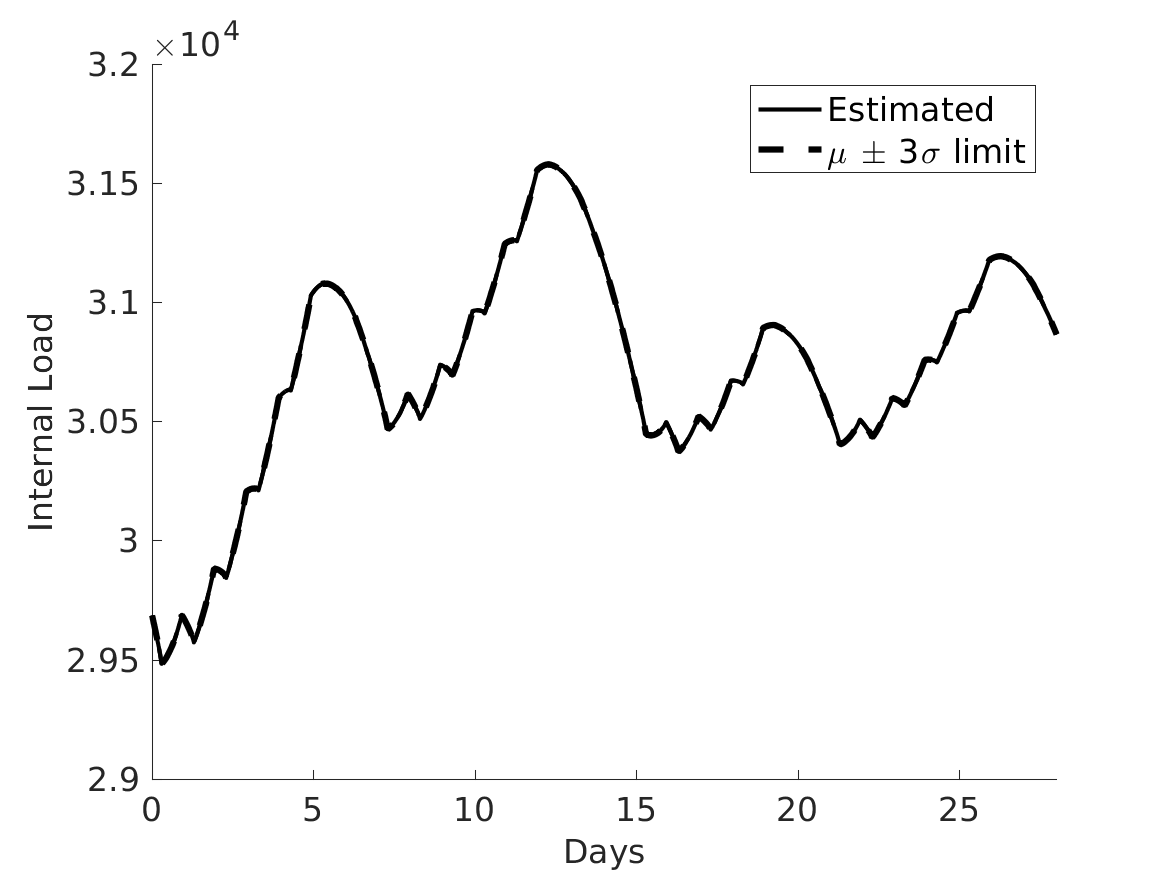}}
\subfigure[Zone 47]{\includegraphics[width=0.45\textwidth]{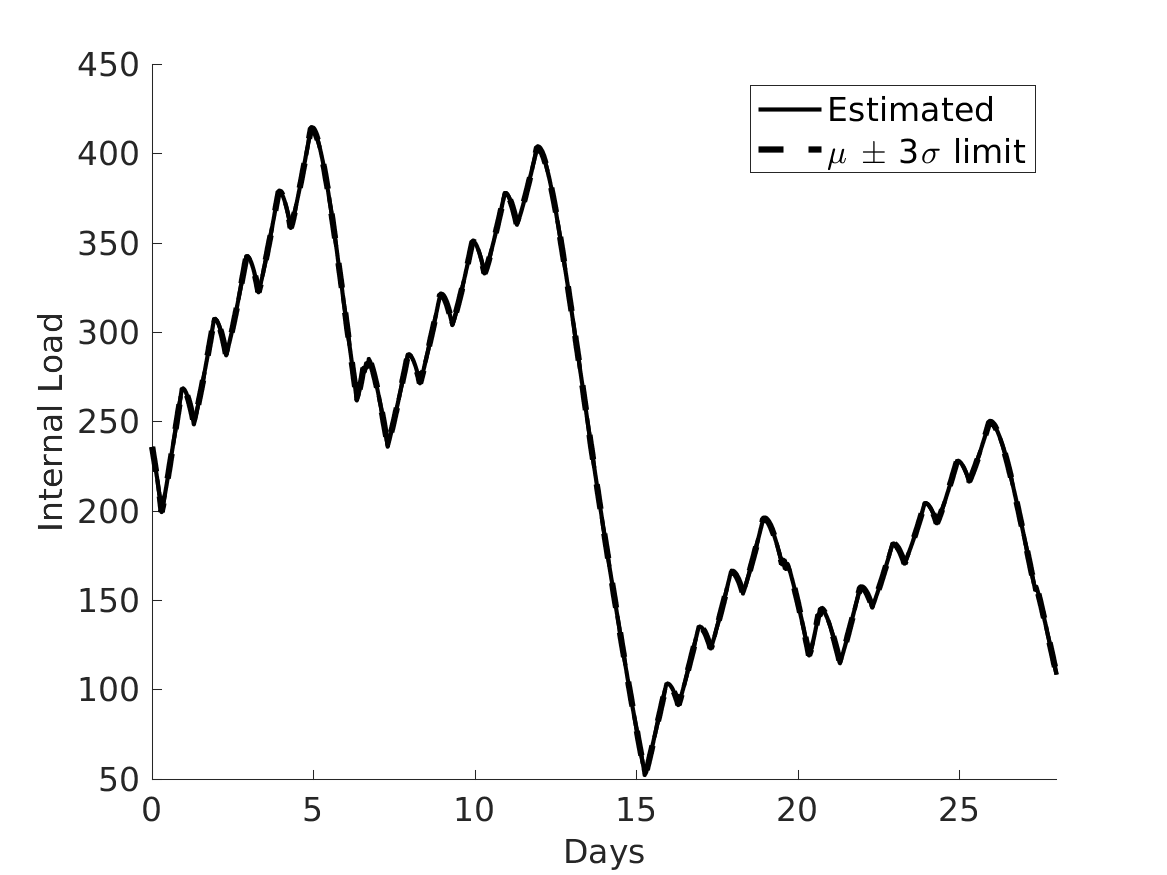}} \\
\subfigure[Zone 80]{\includegraphics[width=0.45\textwidth]{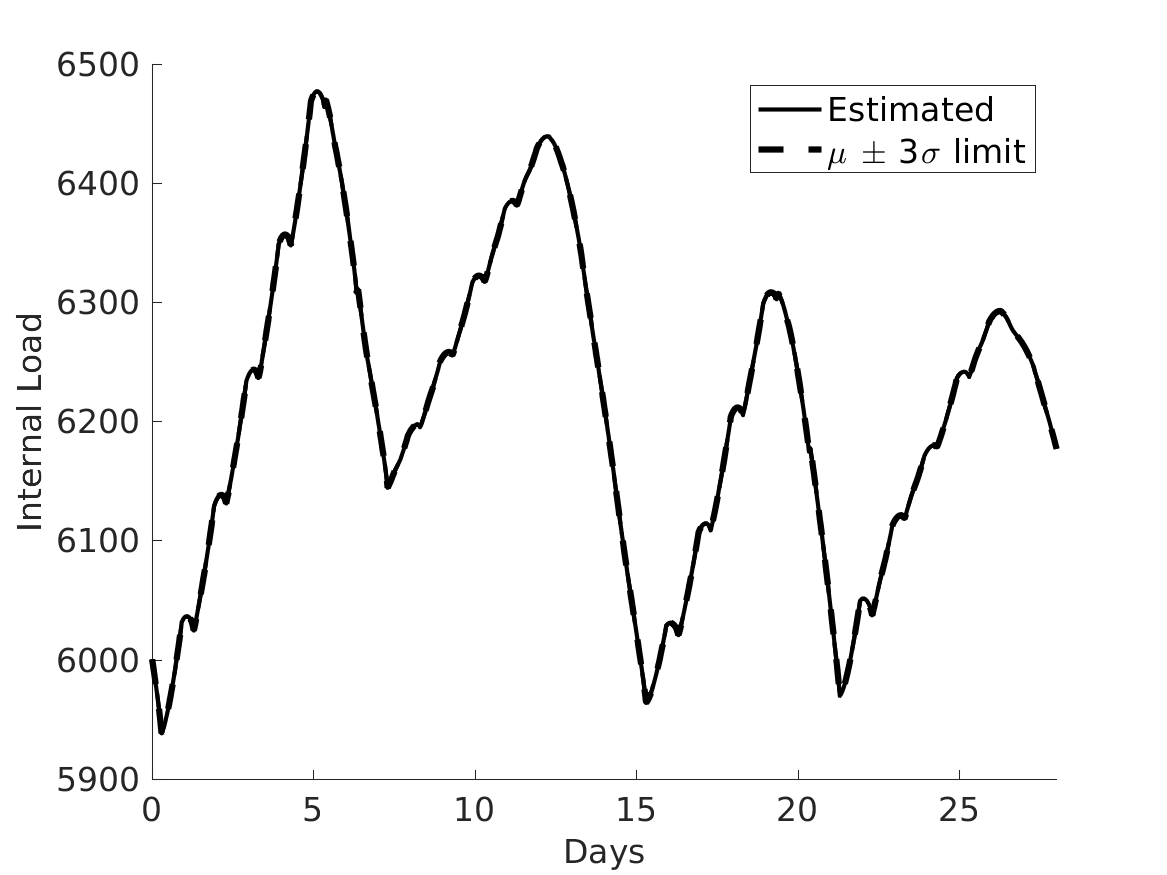}}
\subfigure[Zone 97]{\includegraphics[width=0.45\textwidth]{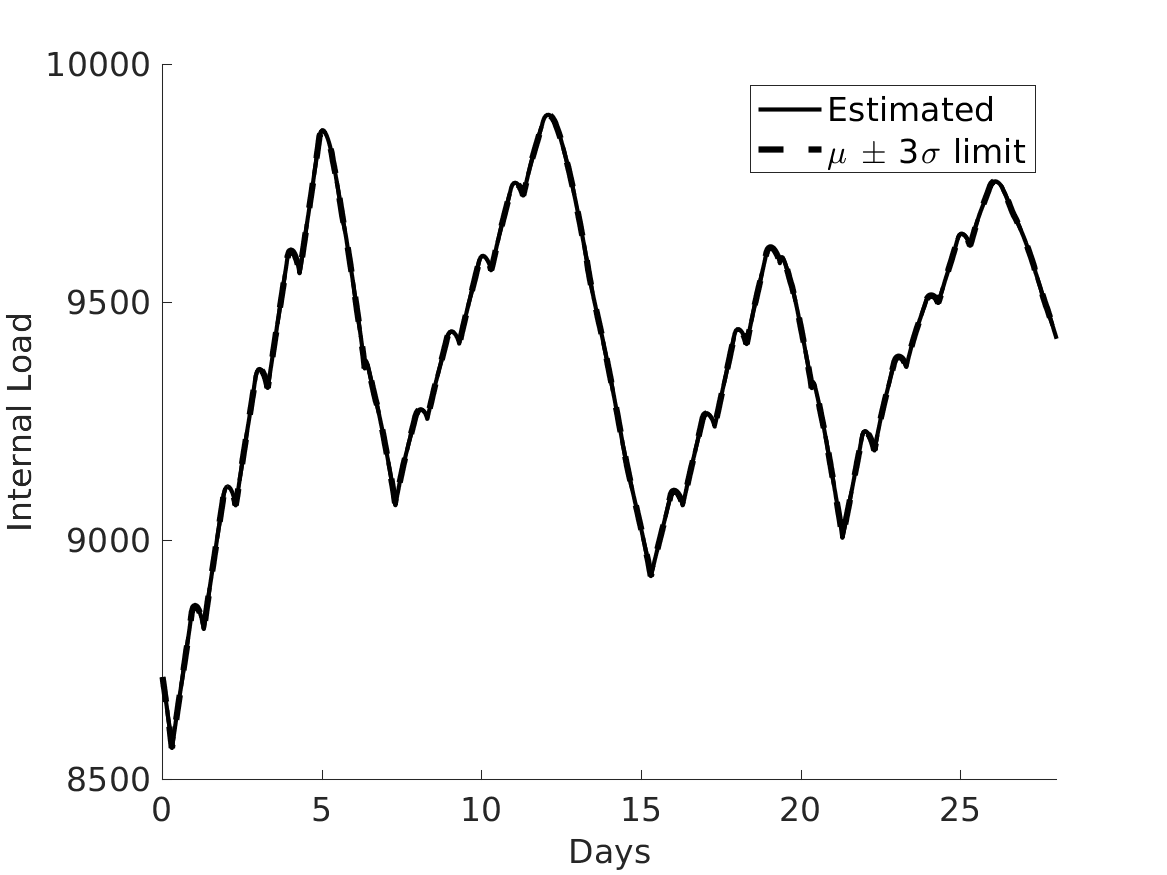}}
\caption{Non-Envelope Loads for (a) Zone 1, (b) Zone 47, (c) Zone 80 and (d) Zone 97 for the month of February}
\label{Zone_Int_temperature_feb}
\end{figure}

\begin{figure}[H]
\centering
\subfigure[Zone 1]{\includegraphics[width=0.45\textwidth]{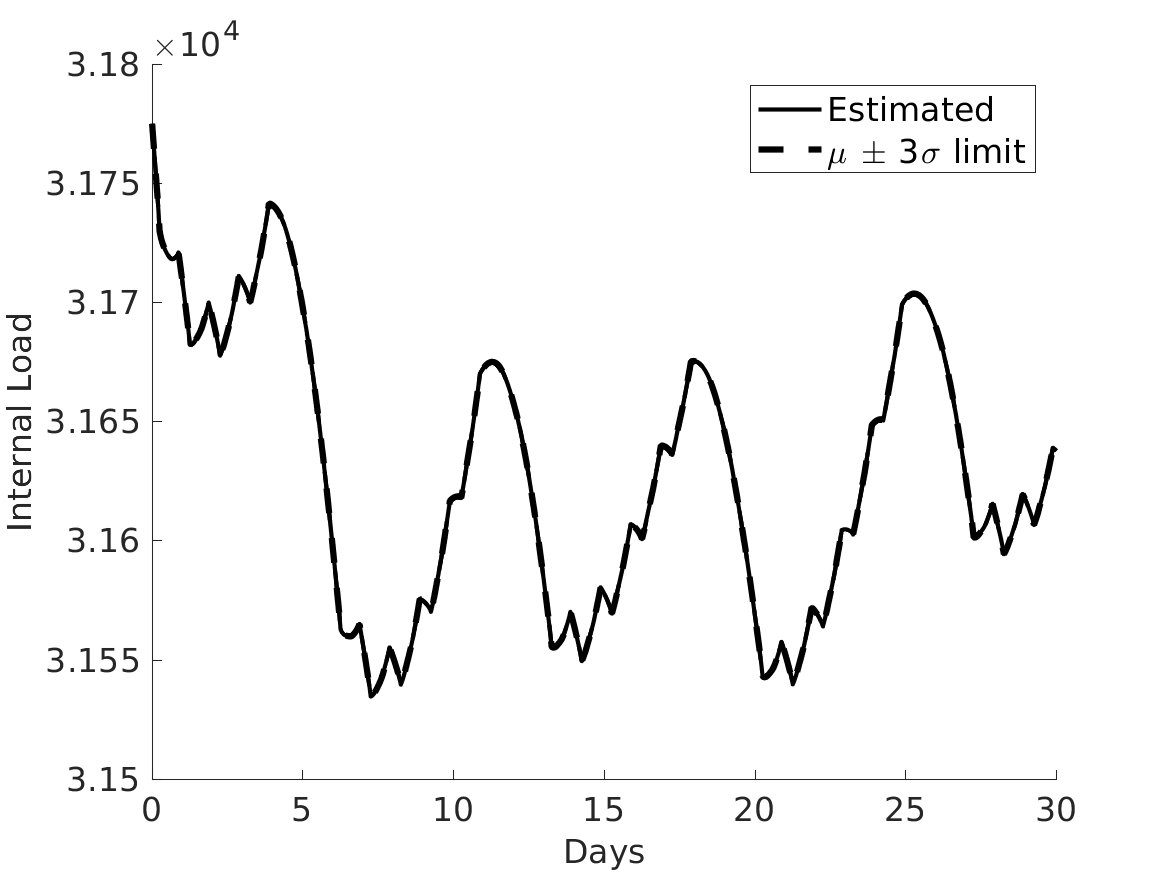}}
\subfigure[Zone 47]{\includegraphics[width=0.45\textwidth]{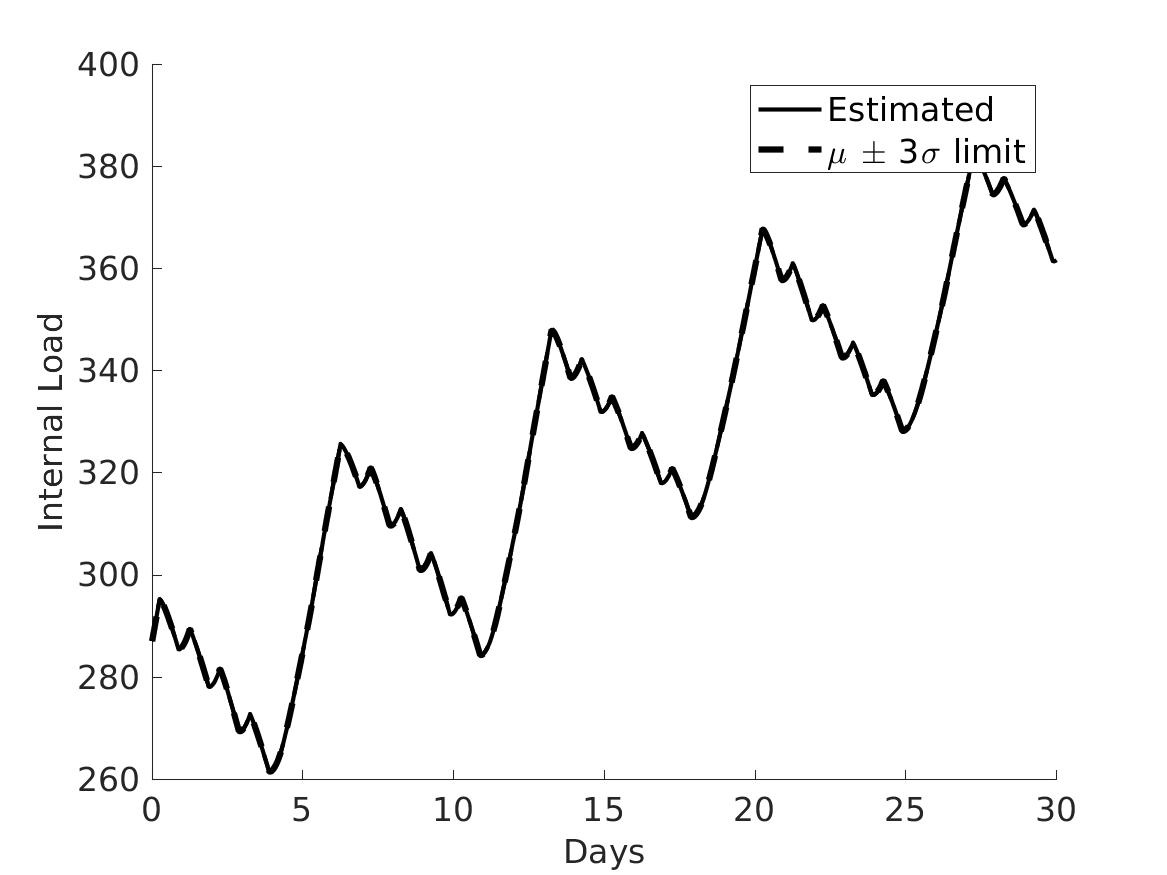}} \\
\subfigure[Zone 80]{\includegraphics[width=0.45\textwidth]{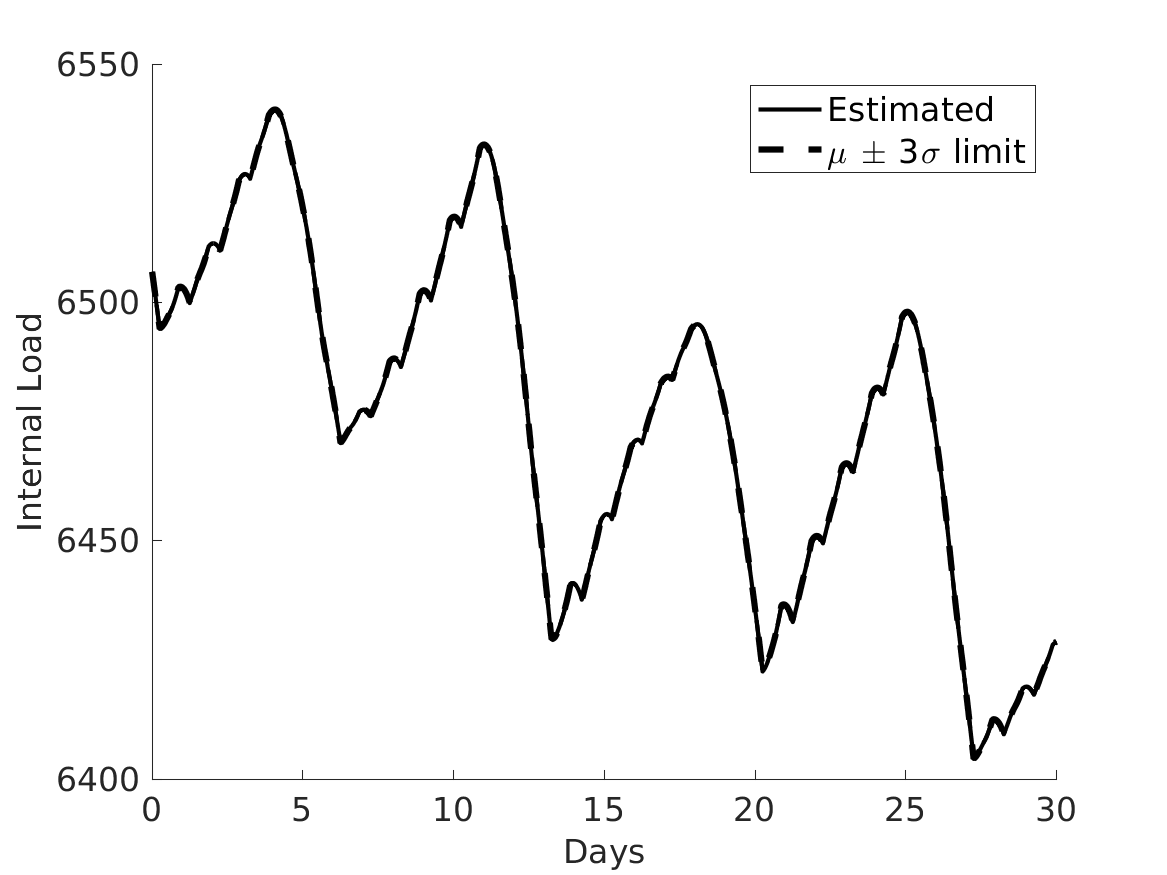}}
\subfigure[Zone 97]{\includegraphics[width=0.45\textwidth]{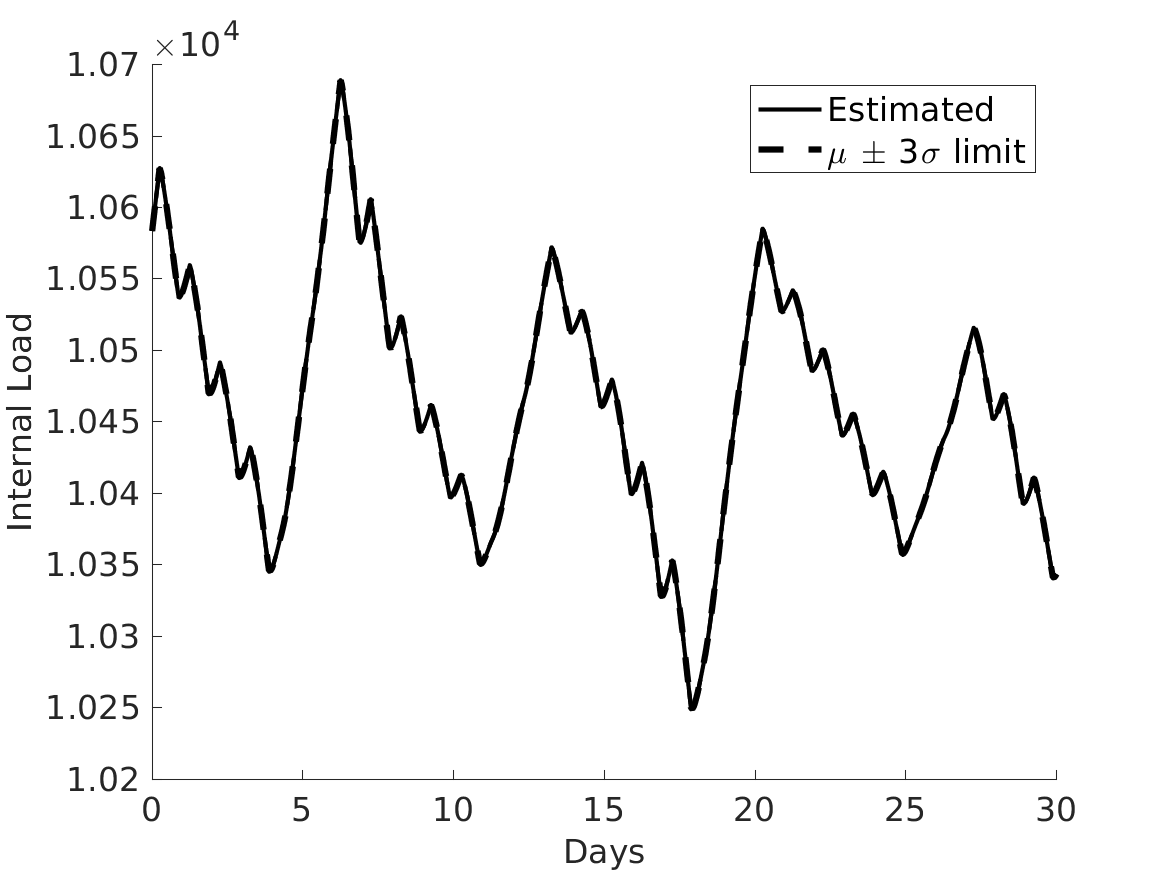}}
\caption{Non-Envelope Loads for (a) Zone 1, (b) Zone 47, (c) Zone 80 and (d) Zone 97 for the month of June}
\label{Zone_Int_temperature_june}
\end{figure}

\begin{figure}[H]
\centering
\subfigure[Zone 1]{\includegraphics[width=0.45\textwidth]{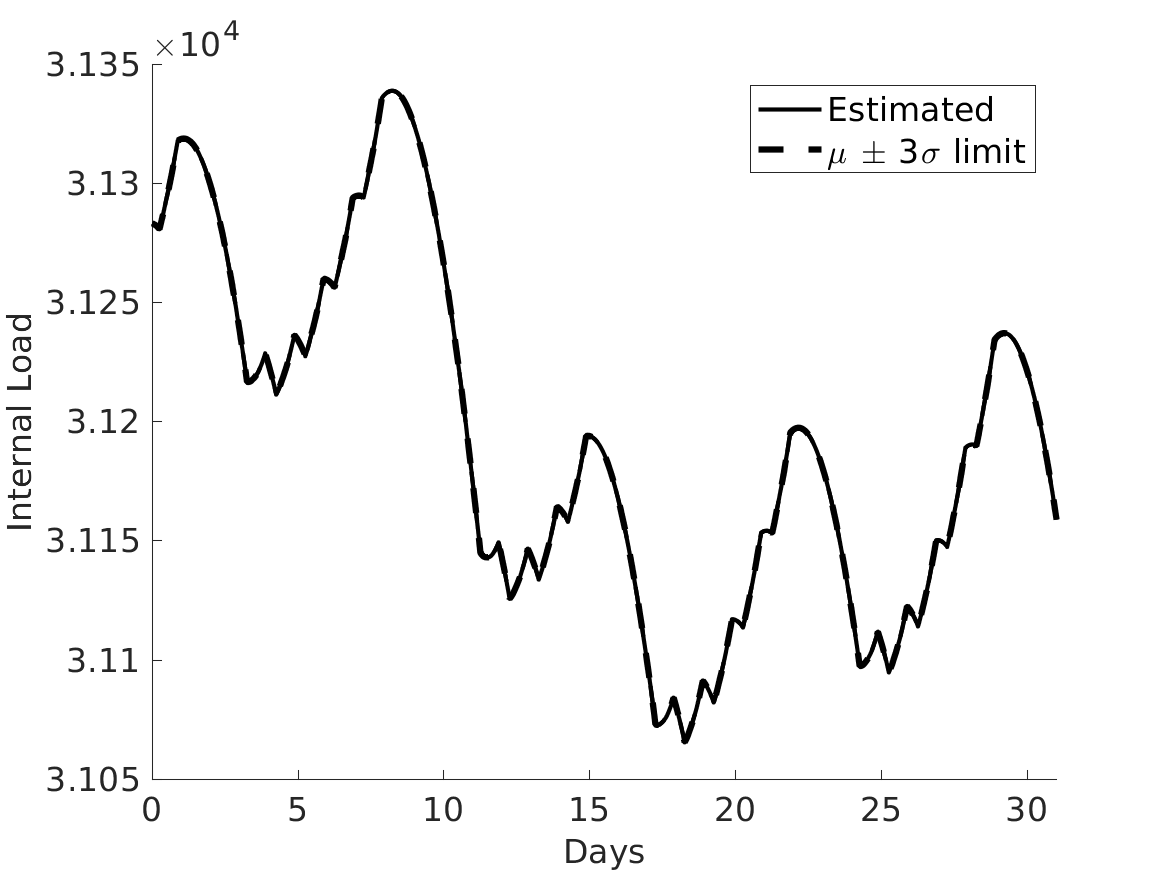}}
\subfigure[Zone 47]{\includegraphics[width=0.45\textwidth]{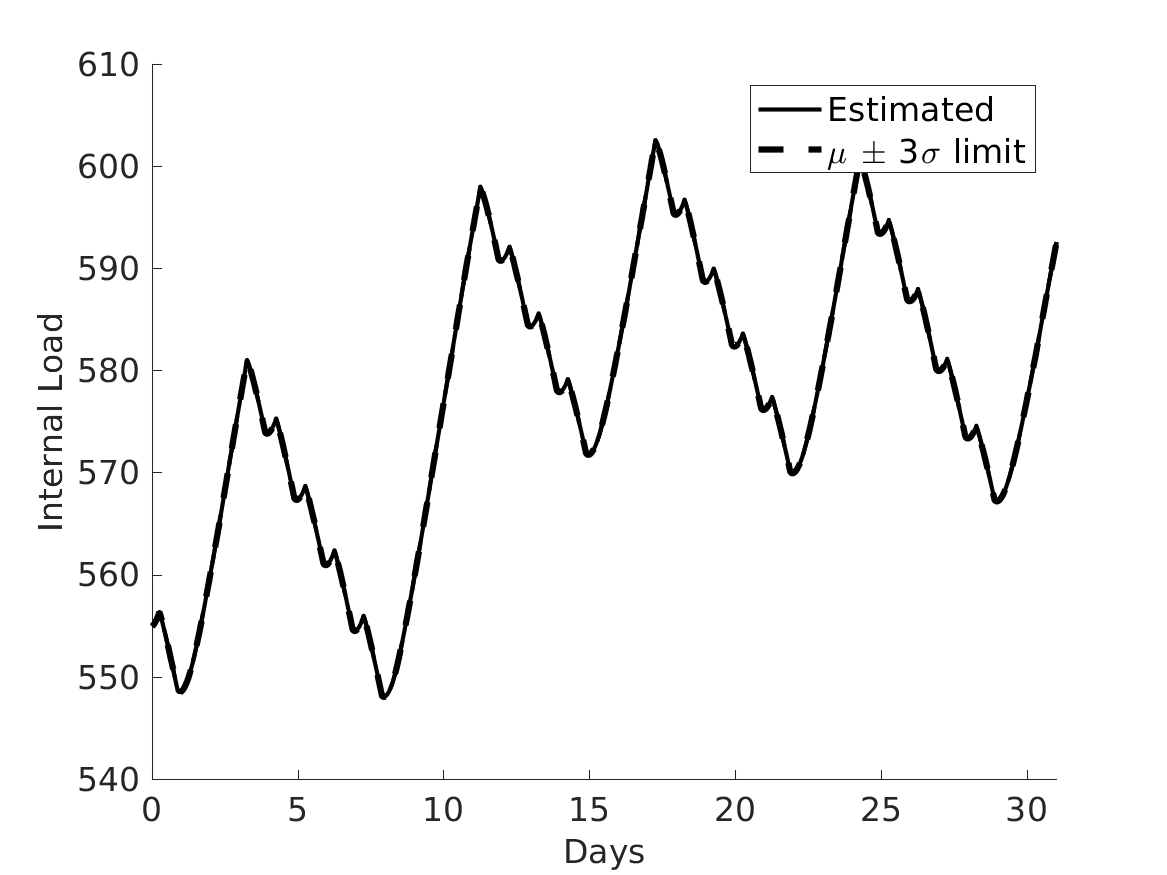}} \\
\subfigure[Zone 80]{\includegraphics[width=0.45\textwidth]{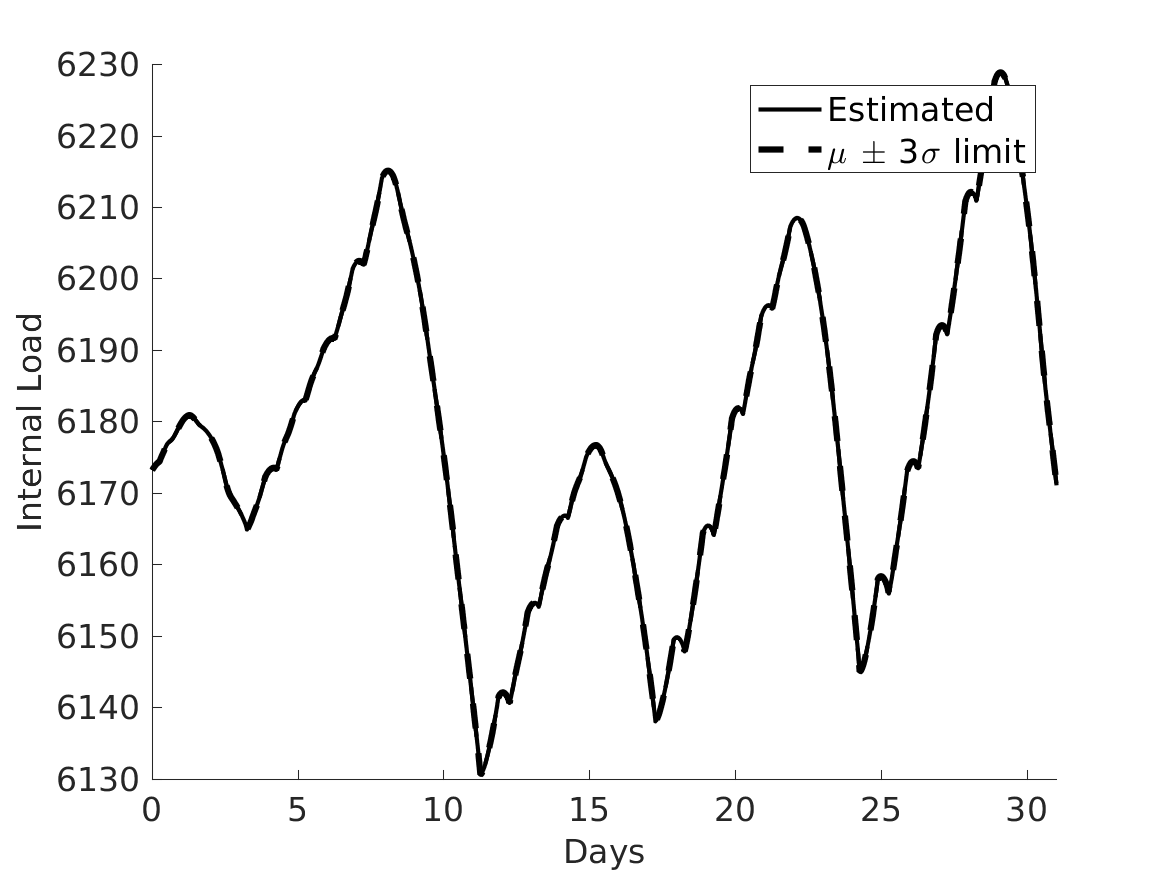}}
\subfigure[Zone 97]{\includegraphics[width=0.45\textwidth]{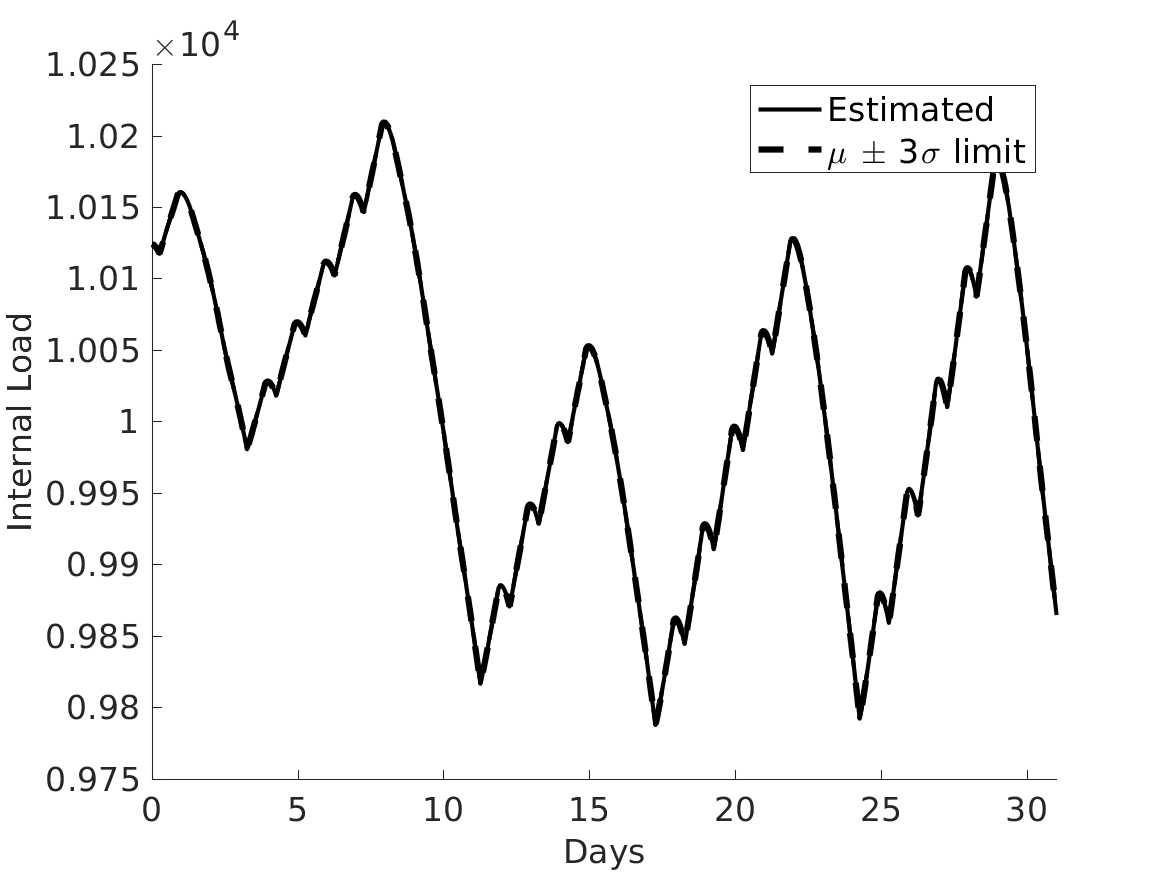}}
\caption{Non-Envelope Loads for (a) Zone 1, (b) Zone 47, (c) Zone 80 and (d) Zone 97 for the month of October}
\label{Zone_Int_temperature_oct}
\end{figure}

Figures~\ref{Zone_Int_temperature_feb},~\ref{Zone_Int_temperature_june}, and~\ref{Zone_Int_temperature_oct} show the estimated non-envelope loads for the four specific zones for the month of February, June, and October respectively. The loads follow the trend of occupancy, as expected, and have daily peaks midday and drop at night.  The thermal load values typically increase daily during weekdays, before decreasing over the weekends.  This is likely due to the high thermal mass associated with the zone, particularly the floor and the exterior walls, which allows the zone to retain heat at night, and provides an increased baseline non-envelope load each day.  Zone 1, for example, has an especially high thermal capacitance due to the surrounding soil, which minimizes the temperature drop in the evenings.         

Because the model has approximately 2700 state variables, and only 132 outputs, the resulting estimations are equally distributed based on the coefficients of the state space coefficients, and the estimated results follow a pattern similar to the estimated zonal temperatures.  The assumption that the thermal loads are slow-changing with a $\dot{T}_k^{int} = 0$ minimizes the estimated nightly reduction in non-envelope load, since it does not permit a drastic change.  Note that these values include the uncertainty due to the baseboard heat.

\begin{figure}[H]
\centering
\subfigure[Zone 47]{\includegraphics[width=0.45\textwidth]{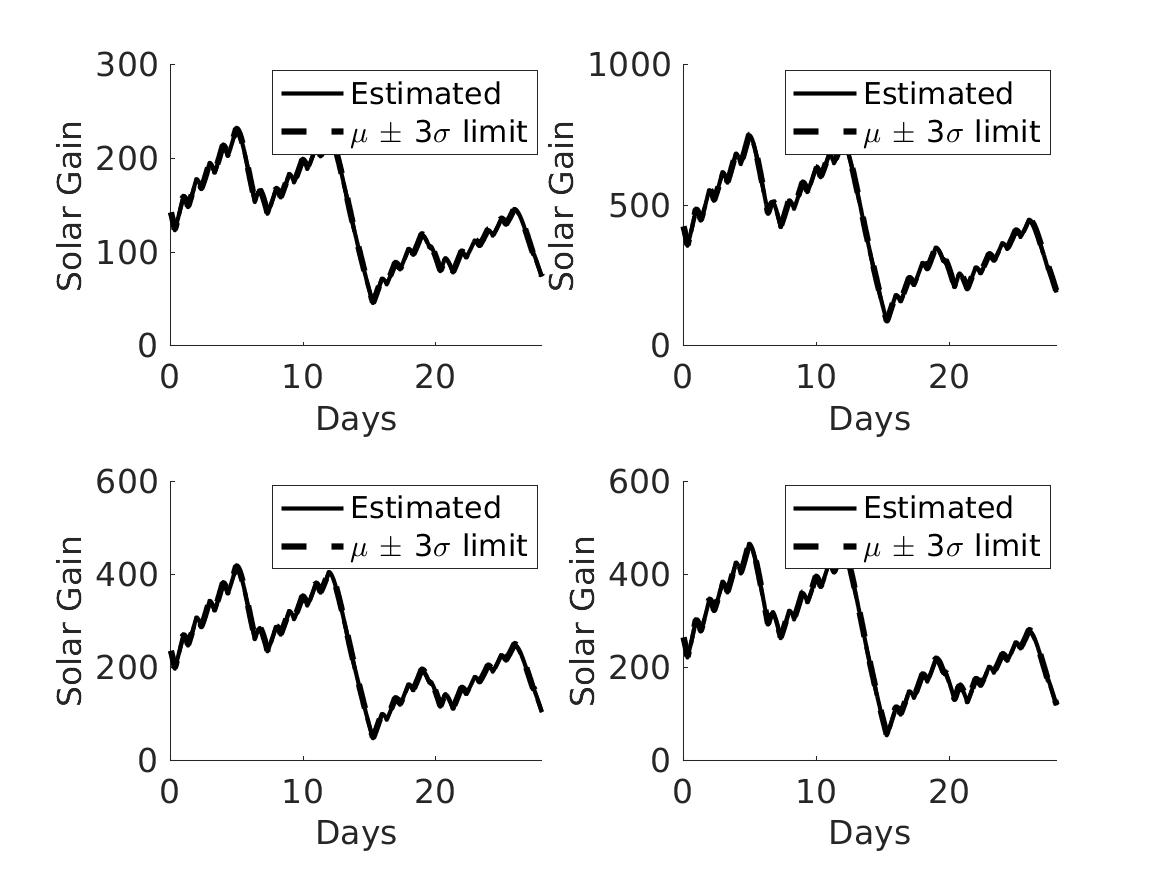}}
\subfigure[Zone 97]{\includegraphics[width=0.45\textwidth]{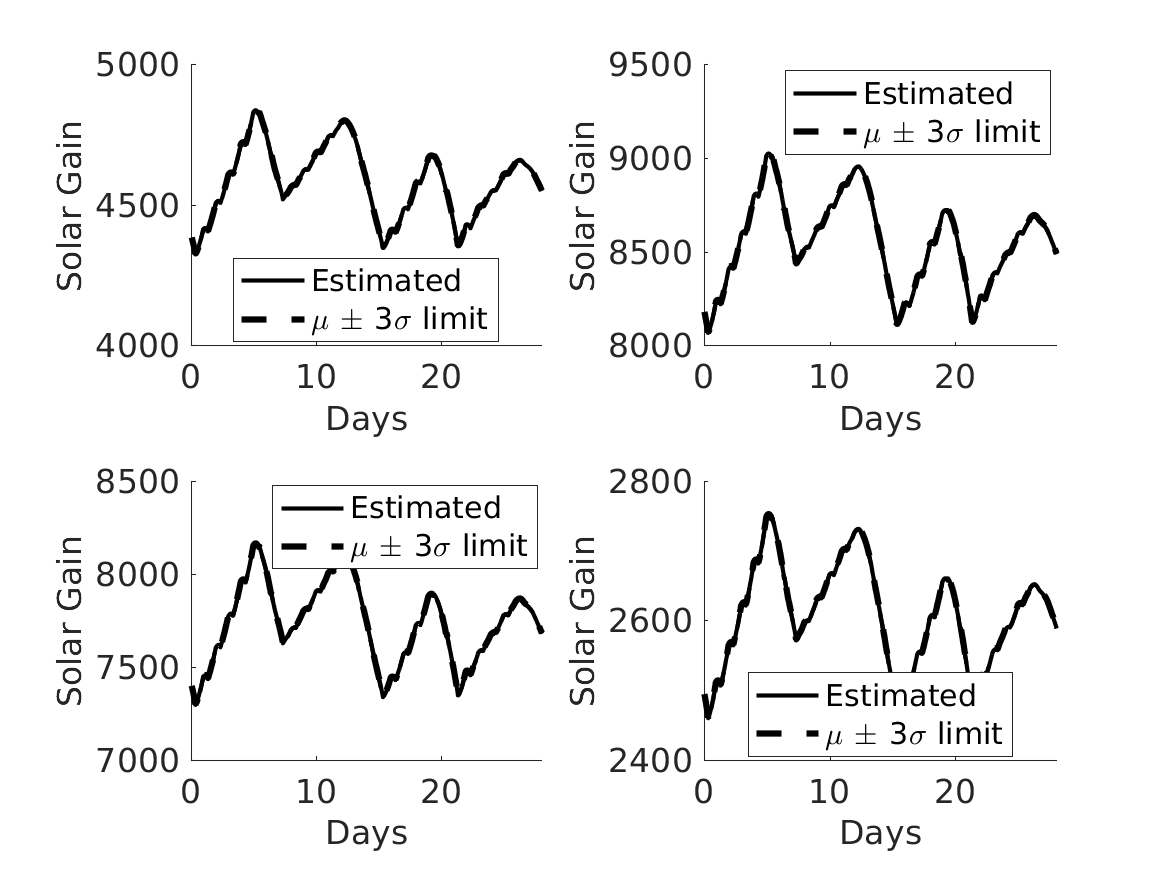}}
\caption{Solar Gains for (a) Zone 47 and (b) Zone 97 for the month of February}
\label{Zone_solar_temperature_feb}
\end{figure}

\begin{figure}[H]
\centering
\subfigure[Zone 47]{\includegraphics[width=0.45\textwidth]{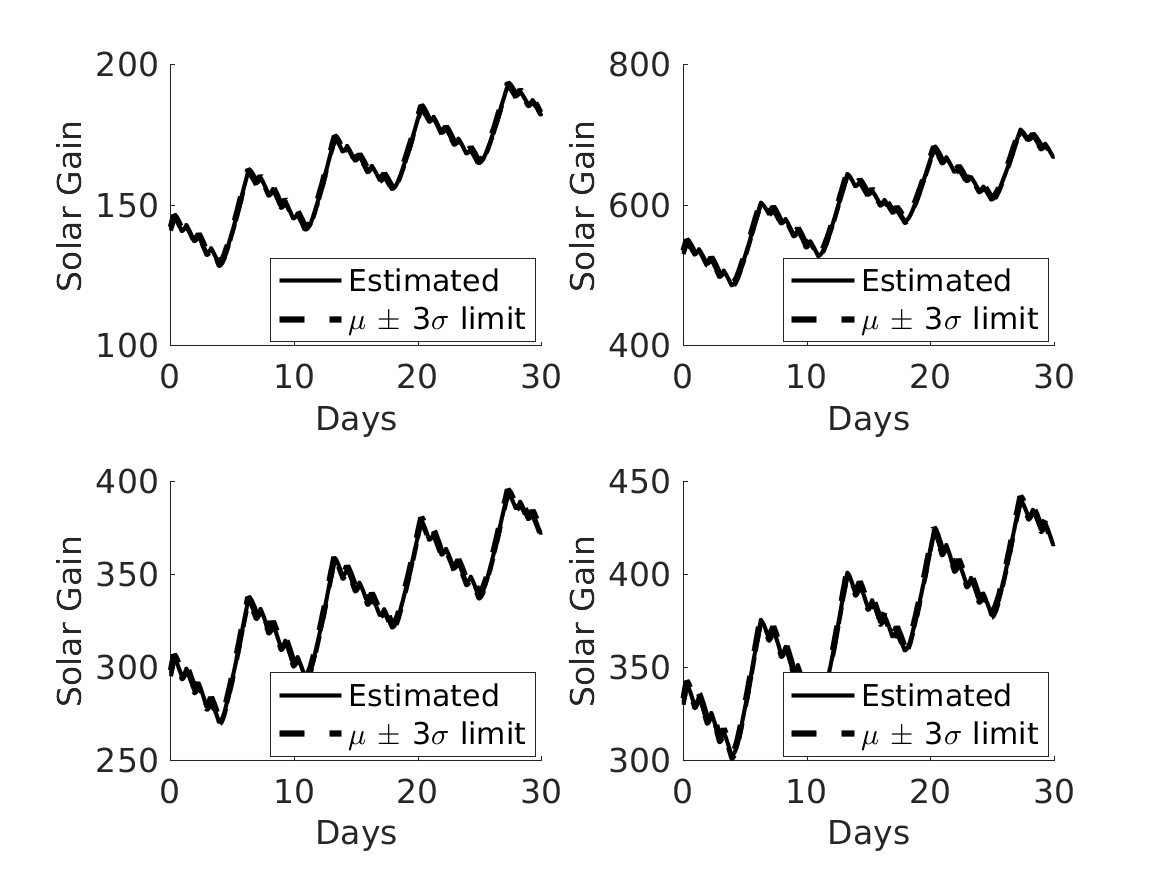}}
\subfigure[Zone 97]{\includegraphics[width=0.45\textwidth]{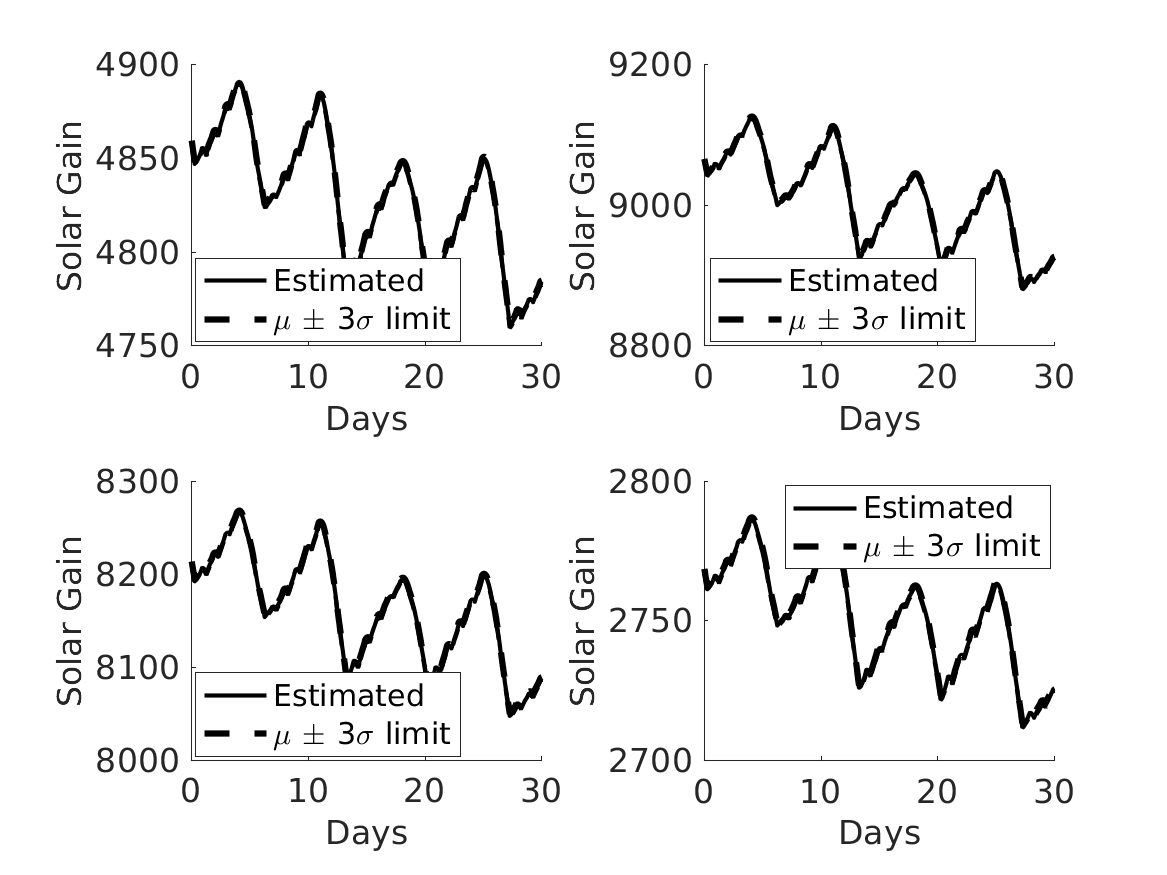}}
\caption{Solar Gains for (a) Zone 47 and (b) Zone 97 for the month of June}
\label{Zone_solar_temperature_june}
\end{figure}

\begin{figure}[H]
\centering
\subfigure[Zone 47]{\includegraphics[width=0.45\textwidth]{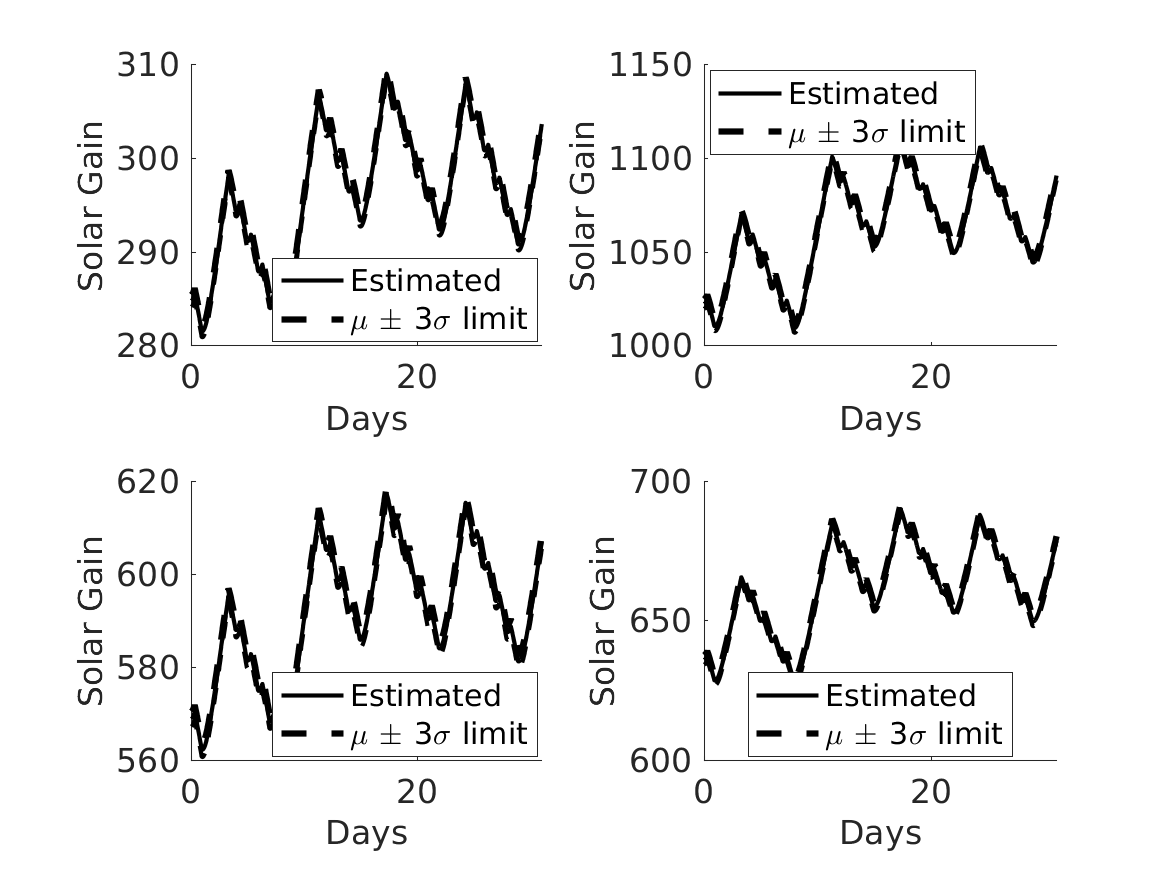}}
\subfigure[Zone 97]{\includegraphics[width=0.45\textwidth]{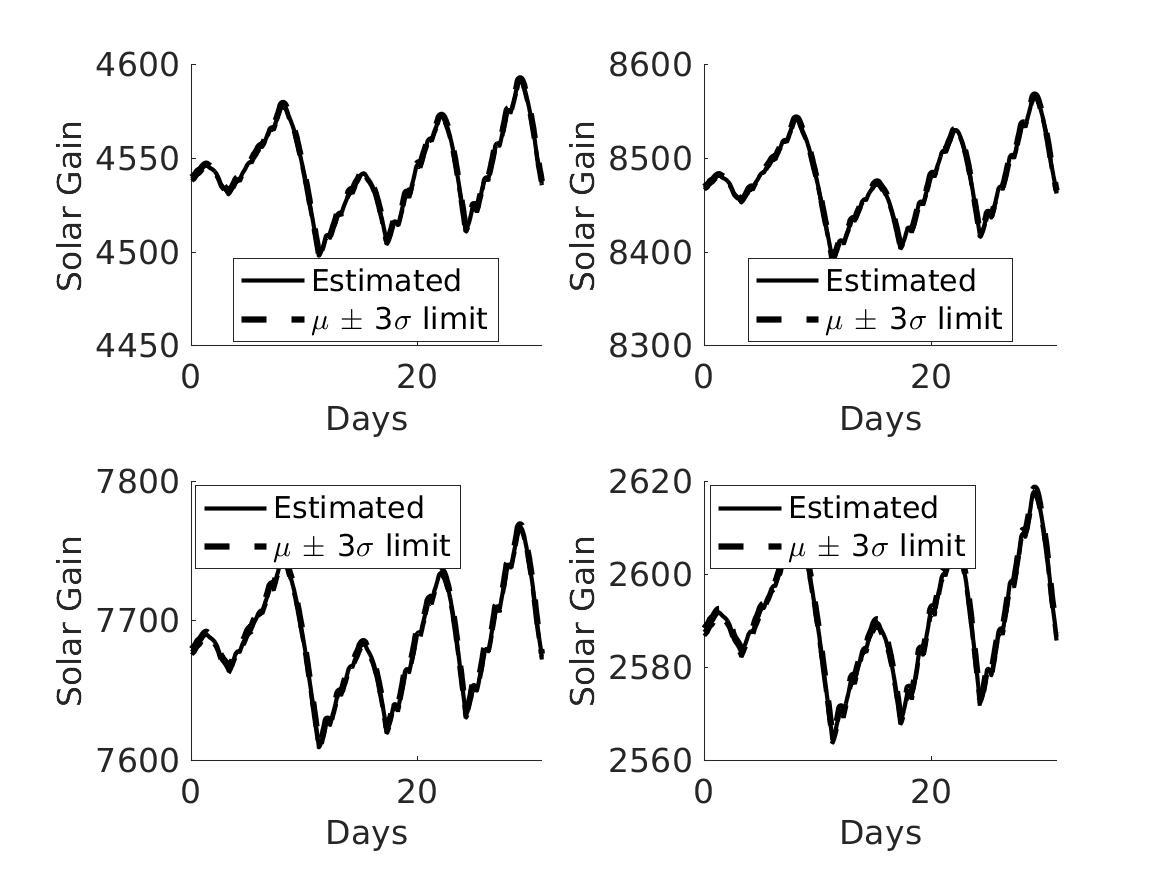}}
\caption{Solar Gains for (a) Zone 47 and (b) Zone 97 for the month of June}
\label{Zone_solar_temperature_oct}
\end{figure}

The solar gains for some of the surfaces for Zone 47 and Zone 97 are displayed in Figure~\ref{Zone_solar_temperature_feb},~\ref{Zone_solar_temperature_june}, and~\ref{Zone_solar_temperature_oct} for the three months (February, June, and October). The graph shows an equal distribution of the solar load into the interior surfaces. This is an expected result, as eQuest models the radiation as uniformly distributed ~\cite{doe2016energyplus}; actual observed results would be likely to differ somewhat ~\cite{he2016simplified} due to the differences in direct and reflected solar radiation on the surfaces.  Note that the surface heat flux somewhat follows the trend of occupancy; this is likely due to the radiant heat output of the lighting in the zone.  This is especially apparent at night, when the solar gain is expected to drop substantially, but some lighting remains on.  As with the thermal loads, the solar gains follow a similar pattern to the estimated outputs due to the high number of estimated state variables.

\section{Conclusion and Future Works}
\label{conclusion}

In this work, a reduced-order thermal BEM has been developed for a large scale office/school building. The lumped capacitance RC network model is used to calculate the non-envelope load and solar gain for interior surfaces. Using the simplified BEM in conjunction with the Weakly Connected Subsystems optimal estimation method allows for estimating these parameters for each thermal zone along with associated uncertainties.

One significant advantage of the methodology is the ability to reduce the computation expense of large-scale dynamical systems while maintaining accuracy and providing uncertainty information at each step.  Thus, the BEM demonstrated in this work can be built upon, adding detail to offer more precision.  For instance, we included infiltration as part of the non-envelope load estimator; instead, this component could be broken out separately to explore the impact of the envelope leakage on the building model. In the same vein, the BEM can be used to compare expected building behavior against the actual building states. This in turn could detect anomalies in the building operation.  Used in conjunction with Model Predictive Control, one can also optimize energy usage and occupant comfort.

The flexibility offered with this technique allows us to vary parameters over time, which is not available in some of the more comprehensive simulation programs.  For example, foam insulation is known to degrade over time, due to the diffusion of thermal gases as the material ages~\cite{de2011longitudinal}.  Most modeling programs, such as eQUEST, require that the thermal conductivity of material remain static over the duration of the simulation.  The conductivity at each time step can be used to provide a more precise representation of the dynamic BEM using the lumped capacitance model.

The technique discussed in this work can be applied to any building and is especially suited for large buildings with diverse occupancy, due to the inter-zone effects, and variable internal loads. With the noise filtered out and the zonal temperatures being estimated, the solar gains and the non-envelope loads can be determined. 

\section*{Acknowledgement}

This material is based upon work supported by the National Science Foundation under Grant NSF CMMI \#1301235. Any opinions, findings, and conclusions or recommendations expressed in this material are those of the author(s) and do not necessarily reflect the views of the National Science Foundation.

\bibliographystyle{ieeetr}

\bibliography{jbs}

\end{document}